\documentclass[useAMS,usenatbib,onecolumn,referee]{mn2e}
\usepackage{amsmath}
\usepackage{amssymb}
\usepackage{graphicx}
\usepackage{euscript}

\newcommand{\beq}{\begin{equation}}
\newcommand{\eeq}{\end{equation}}

\newcommand{\bfb}{\mbox{\boldmath $b$}}

\newcommand{\bfh}{\mbox{\boldmath $h$}}
\newcommand{\bfk}{\mbox{\boldmath $k$}}

\newcommand{\bfv}{\mbox{\boldmath $v$}}
\newcommand{\bfx}{\mbox{\boldmath $x$}}

\newcommand{\bfr}{\mbox{\boldmath $r$}}
\newcommand{\bfA}{\mbox{\boldmath $A$}}
\newcommand{\bfB}{\mbox{\boldmath $B$}}

\newcommand{\bfE}{\mbox{\boldmath $E$}}
\newcommand{\bfF}{\mbox{\boldmath $F$}}
\newcommand{\bfH}{\mbox{\boldmath $H$}}
\newcommand{\bfK}{\mbox{\boldmath $K$}}
\newcommand{\bfM}{\mbox{\boldmath $M$}}
\newcommand{\bfR}{\mbox{\boldmath $R$}}
\newcommand{\bfU}{\mbox{\boldmath $U$}}
\newcommand{\bfV}{\mbox{\boldmath $V$}}

\newcommand{\bfX}{\mbox{\boldmath $X$}}

\newcommand{\bfxi}{\mbox{\boldmath $\xi$}}
\newcommand{\ex}{\mbox{{\boldmath $e$}}_{1}}
\newcommand{\ey}{\mbox{{\boldmath $e$}}_{2}}
\newcommand{\ez}{\mbox{{\boldmath $e$}}_{3}}

\newcommand{\bfemf}{\mbox{\boldmath ${\cal E}$}}
\newcommand{\bnabla}{\mbox{\boldmath $\nabla$}}
\newcommand{\cross}{\mbox{\boldmath $\times$}}
\newcommand{\cendot}{\mbox{\boldmath $\cdot\,$}}

\begin{document}

\title[Stochastic alpha shear dynamo]
{Large--scale dynamo action due to $\alpha$ fluctuations in a linear 
shear flow}

\author[Sridhar \& Singh]{S. Sridhar$^{1,4}$ and Nishant K. Singh$^{2,3,5}$\\
 $^{1}$Raman Research Institute, Sadashivanagar, Bangalore 560 080, India\\
 $^{2}$Inter--University Centre for Astronomy and Astrophysics, Post Bag 4,
Ganeshkhind, Pune 411 007, India\\
 $^{3}$Nordita, KTH Royal Institute of Technology and Stockholm University,
Roslagstullsbacken 23, SE-10691 Stockholm, Sweden\\
 $^{4}$ssridhar@rri.res.in\,, 
 $^{5}$nishant@nordita.org
}

\pagerange{\pageref{firstpage}--\pageref{lastpage}} \pubyear{2013}

\maketitle

\label{firstpage}

\begin{abstract}
We present a model of large--scale dynamo action in a shear flow that has stochastic, zero--mean fluctuations of the $\alpha$ parameter. This is based on a minimal extension of the Kraichnan--Moffatt model, to include a background linear shear and Galilean--invariant $\alpha$--statistics. Using the first order smoothing approximation we derive a linear integro--differential equation for the large--scale magnetic field, which is non perturbative in the shearing rate $S\,$, and the $\alpha$--correlation time
$\tau_\alpha\,$. The white--noise case, $\tau_\alpha = 0\,$, is solved exactly, and it is concluded that the necessary condition for dynamo action is identical to the Kraichnan--Moffatt model without shear; this is because white--noise does not allow for memory effects, whereas shear needs time to act. To explore memory effects we reduce the integro--differential equation to a partial differential equation, valid for slowly varying fields when 
$\tau_\alpha$ is small but non zero. Seeking exponential modal solutions, we solve the modal dispersion relation and obtain an explicit expression for the growth rate as a function of the six independent parameters of the problem. A non zero $\tau_\alpha$ gives rise to new physical scales, and dynamo action is completely different from the white--noise case; e.g. 
even weak $\alpha$ fluctuations can give rise to a dynamo. We argue that, 
at any wavenumber, both Moffatt drift and Shear always contribute to increasing the growth rate. Two examples are presented: (a) a Moffatt drift dynamo in the absence of shear; (b) a Shear dynamo in the absence of Moffatt drift. 
\end{abstract}


\section{Introduction}
The magnetic fields observed in many astrophysical bodies --- such as the Sun, stars, galaxies and clusters of galaxies --- are thought to be 
generated by electric currents in the turbulent plasmas flows in these
objects \citep{Mof78, Par79, KR80,RSS98,Kul04,BS05}. The field shows structure on a wide range of scales, both smaller and larger than that 
of the underlying turbulence. Of particular interest to the present investigation is the large--scale ordered structure of the field. When the turbulent motions are helical (i.e when mirror--symmetry is broken), the well--known $\alpha$--effect can amplify seed magnetic fields and maintain them in the face of turbulent dissipation \citep{Mof78, Par79, KR80}. However, it is not clear whether astrophysical turbulence has a mean helicity that is large enough to sustain such a large--scale turbulent dynamo. A natural question arises: is there dynamo action for turbulence whose helicity vanishes (not instantaneously, but) on average? \citet{Kra76} considered dynamo action when $\alpha$ is a stochastic quantity with zero mean, and demonstrated that the $\alpha$ fluctuations would contribute to a decrease in the turbulent diffusivity. When the fluctuations are large enough, there is negative diffusion, and the magnetic field grows on all spatial scales, with the fastest rates of growth on the smallest scales. The Kraichnan model was generalized by \citet{Mof78}, to include a statistical correlation between the fluctuations of $\alpha$ and its spatial gradient. In the simplest 
case this correlation contributes a constant drift velocity to the dynamo equation, and does not influence dynamo action. A feature common to astrophysical flows is velocity shear, and it is only relatively recently that the question of large--scale dynamo action in shear flows with $\alpha$ fluctuations began receiving attention \citep{VB97,Sok97,Sil00}. The interest in this question has grown significantly, and a number of investigations have pursued this problem both analytically and 
numerically \citep{Pro07,BRRK08,You08,KR08,RK08,SurSub09,HMS11,McW12,MB12,Pro12,RP12,TC13}.
The numerical simulations of \citet{You08} and \citet{BRRK08} are  demonstrations of large--scale dynamo action in a shear flow with 
turbulence that is, on average, non--helical. The evidence for the growth of a large--scale magnetic field seems compelling, as can be seen from the `butterfly diagrams' of Figure~3 in \citet{You08} and Figure~8 in \citet{BRRK08}. It is also significant, as shown in Figure~10 of \citet{BRRK08}, that the $\alpha$ parameter fluctuates about a mean value close to zero. 

The goal of this paper is to study large--scale dynamo action by 
extending the Kraichnan--Moffatt (KM) model in a minimal manner, 
in order to include a background linear shear flow. This is done in \S~2 where the $\alpha$ fluctuations are required to respect Galilean invariance, a symmetry natural to linear shear flows. The integro--differential equation governing the evolution of the large--scale magnetic field is derived in \S~3, using the first order smoothing approximation, shearing coordinates  and Fourier representation similar to \cite{SS09a,SS09b,SS10,SS11}. This equation is non perturbative in the shearing rate $S$, and the $\alpha$--correlation time $\tau_\alpha\,$. With no further approximation, this is applied in \S~4 to the case of white--noise 
$\alpha$ fluctuations (for which $\tau_\alpha = 0\,$) and general conclusions are derived regarding the limited nature of dynamo action. In \S~5 the integro--differential equation is simplified to an 
ordinary differential equation, when $\tau_\alpha \neq 0$ is small and the large--scale magnetic field is slowly varying. This is used to explore
one--dimensional propagating waves in \S~6: the dispersion relation is solved and an explicit expression for the growth rate function is obtained. In \S~7 we present an interpolation formula for the growth rate and study dynamo action due to Kraichnan diffusivity, Moffatt drift and shear. We conclude in \S~8.

\section{Definition of the model}
 
\subsection{Outline of the Kraichnan--Moffatt (KM) model}
\label{KMmodel}

Following \cite{Kra76} we model small--scale turbulence in the absence of mean motions as random velocity fields, $\{\bfv(\bfX, \tau)\}$, where $\bfX = (X_1, X_2, X_3)$ is the position vector with components given in a fixed orthonormal frame $(\ex, \ey, \ez)$, and $\tau$ is the time variable. The ensemble is assumed to have \emph{zero mean isotropic velocity fluctuations, with uniform and constant kinetic energy density per unit mass, and slow helicity fluctuations}. Zero mean isotropic velocity fluctuations implies that:
\beq
\left\langle v_i\right\rangle \;=\; 0\,;\qquad
\left\langle v_i v_j\right\rangle \;=\;\delta_{ij} v_0^2 \;;\qquad
\left\langle v_i \frac{\partial v_j}{\partial X_n}\right\rangle \;=\;
\epsilon_{inj}\, \mu(\bfX, \tau)\,,
\eeq

\noindent
where $v_0^2 = \left\langle v^2/3\right\rangle =$~two--thirds of the mean kinetic
energy density per unit mass, and $\mu(\bfX, \tau) = \left\langle\bfv\cendot(\bnabla\cross\bfv)\right\rangle/6 =$~one--sixth of the helicity 
density. Uniform and constant kinetic energy density per unit mass means that $v_0^2$ is a constant number across the ensemble. Let $\ell_0$ 
be the eddy size and $\tau_c$ be the velocity correlation time. The correlation time need not equal the `eddy turnover time' $\tau_0=\ell_0/v_0\,$: for instance, $\tau_c\sim\tau_0$ for fully developed turbulence, whereas $\tau_c\ll\tau_0$ for a $\bfv\,$ which is close to white--noise. By slow helicity fluctuations we mean that the spatial and temporal scales of variation of $\mu(\bfX, \tau)$ are assumed to be much larger than $\ell_0$ and $\tau_c$. 

When electric currents in the fluid give rise to a magnetic field, the 
action of the stochastic velocity field makes the magnetic field stochastic
as well. Let $\bfB(\bfX, \tau)$ be the \emph{meso--scale magnetic field}, 
obtained by averaging over the above ensemble.  Then the space--time evolution of $\bfB(\bfX, \tau)$ is given by the \emph{dynamo equation} \citep{Mof78,KR80,BS05}:
\beq
\frac{\partial \bfB}{\partial \tau} \;=\; \bnabla\cross\left[\,\alpha(\bfX, \tau) \bfB\,\right]\,+\,\eta_T\bnabla^2\bfB
\,;\qquad\qquad \bnabla\cendot\bfB \;=\; 0\,,
\label{DynEqn}
\eeq

\noindent
where
\begin{eqnarray}
\alpha &\;=\;& -2\tau_c\mu(\bfX, \tau)\,,\qquad\qquad
\eta_T \;=\; \eta \,+\,\eta_t \;=\; \mbox{total diffusivity}\,,
\nonumber\\[1ex]
\eta &\;=\;& \mbox{microscopic diffusivity}\,,\qquad\qquad
\eta_t \;=\; \tau_c v_0^2 \;=\; \mbox{turbulent diffusivity}\,.
\label{pardef}
\end{eqnarray}

\noindent
The above expressions for the turbulent transport coefficients $\alpha$ and
$\eta_t$ are valid in the high--conductivity limit, under the first order smoothing approximation (FOSA), assuming isotropic background turbulence.
The next step is to consider $\alpha(\bfX, \tau)$ itself to be  
drawn from a \emph{superensemble}, while $\,\eta_T$ is kept constant.
It is assumed that $\alpha(\bfX, \tau)$ is a \emph{statistically stationary, homogeneous,
random function of $\bfX$ and $\tau$ with zero mean, $\overline{\alpha(\bfX, \tau)}=0\,$, and two--point space--time correlation function}:
\begin{eqnarray}
\overline{\alpha(\bfX, \tau) \alpha(\bfX', \tau')} &\;=\;& 
2{\cal A}(\bfX-\bfX')\,{\cal D}(\tau-\tau')\,,\qquad\mbox{with}
\nonumber\\[1em]
2\int_0^{\infty}\,{\cal D}(\tau){\rm d}\tau &\;=\;& 1\,,\qquad\qquad
{\cal A}({\bf 0}) \;=\; \eta_{\alpha} \,\geq\, 0\,.
\label{KMcorr}
\end{eqnarray} 

\noindent
Here $\eta_{\alpha}$ is the \emph{$\alpha$--diffusivity}, introduced first by \citet{Kra76}. The meso--scale field is split as $\bfB=\overline{\bfB}+\bfb$, where $\overline{\bfB}$ is the \emph{large--scale magnetic field} which is equal to the superensemble--average of the meso--scale field, 
and $\bfb$ is the \emph{fluctuating magnetic field}. An 
equation for $\overline{\bfB}$ is derived in \S~7.11 of \citet{Mof78}, 
using FOSA for the fluctuating field:
\beq
\frac{\partial \overline{\bfB}}{\partial \tau} \;=\; \bnabla\cross\left[\,
\bfV_{\!\!M}\cross\overline{\bfB}\,\right] \,+\, \eta_K\bnabla^2\overline{\bfB}\,;\qquad \bnabla\cendot\overline{\bfB} = 0\,,
\label{KMmfe}
\eeq

\noindent
where 
\begin{eqnarray}
\eta_K &\;=\;& \eta_T \,-\, \eta_{\alpha} \;=\; \mbox{Kraichnan diffusivity}\,,
\nonumber\\[1em] 
\bfV_{\!\!M} &\;=\;& -\left(\frac{\partial {\cal A}(\bfxi)}{\partial\bfxi}\right)_{\bfxi =\bf0} \;=\; 
\int_0^\infty
\overline{\alpha(\bfX, \tau)\bnabla\!\alpha(\bfX, 0)}\,{\rm d}\tau 
\;=\; \mbox{Moffatt drift velocity}\,, 
\label{kmcondef}
\end{eqnarray}

\noindent
are the two new constants that determine the behaviour of the large--scale magnetic field. Note that the $\alpha$ diffusivity contributes a decrement to the diffusivity, and hence aids dynamo action.
The general solution to (\ref{KMmfe}) in an infinitely extended medium is:
\beq
\overline{\bfB}(\bfX, \tau) \;=\; \int\frac{{\rm d}^3K}{(2\pi)^3}
\exp{\!\left[\mathrm{i}\bfK\cendot\bfX\right]}\,\widetilde{\overline{\bfB}}(\bfK, \tau)\,,
\eeq

\noindent
where 
\beq
\widetilde{\overline{\bfB}}(\bfK, \tau) \;=\; \widetilde{\overline{\bfB}}_0(\bfK)\exp{\!\left[-\eta_K K^2\tau \,-\, \mathrm{i}\bfK\cendot\bfV_{\!\!M}\tau\right]}\,;\qquad \bfK\cendot \widetilde{\overline{\bfB}}_0 = 0\,.
\eeq

\noindent
\emph{Thus the large--scale magnetic field is a linear superposition of
transverse waves that translate uniformly with velocity $\bfV_{\!\!M}\,$,
while their amplitudes grow or decay according to whether $\eta_K$ is negative or positive}.
The growth/decay rates are proportional to $K^2$, so the fastest growing/decaying modes
are those with the smallest spatial scales $\sim \mbox{few}\;\ell_0\,$.

\subsection{The Kraichnan--Moffatt model with shear}
\label{KMextn}

In the presence of a background linear shear flow with velocity field $\bfV(\bfX)$, the dynamo equation~(\ref{DynEqn}) acquires an additional term, 
$\bnabla\cross\left[\bfV\cross\bfB\right]$, on the
right hand side. For $\bfV=SX_1\ey$, the dynamo equation is:
\beq
\left(\frac{\partial}{\partial\tau} \,+\, SX_1\frac{\partial}{\partial X_2}\right)\!\bfB\;-\; SB_1\ey \;=\; \bnabla\cross\left[\,\alpha(\bfX, \tau) \bfB\,\right]\,+\,\eta_T\bnabla^2\bfB\,,
\qquad\qquad \bnabla\cendot\bfB \;=\; 0\,. 
\label{totmf_ensavd_expl}
\eeq

\noindent
Splitting $\bfB=\overline{\bfB}+\bfb$, and Reynolds averaging equation~(\ref{totmf_ensavd_expl}), we find that the evolution of the large--scale field 
is governed by:
\begin{eqnarray}
\left(\frac{\partial}{\partial\tau} \,+\, SX_1\frac{\partial}{\partial X_2}\right)\overline{\bfB} \;-\; S\overline{B_1}\ey 
&\;=\;& \bnabla\cross\overline{\bfemf} \;+\; \eta_T\bnabla^2\overline{\bfB}\,,\qquad\qquad \bnabla\cendot\overline{\bfB} \;=\; 0\,,
\label{meanalpfluc}\\[2ex]
\mbox{where}\qquad\overline{\bfemf} &\;=\;& \overline{\alpha(\bfX, \tau)\bfb(\bfX, \tau)}\,.
\label{meanEMF}
\end{eqnarray}

\noindent
To calculate $\overline{\bfemf}$, the \emph{electromotive force (EMF)}, 
we need to solve for the fluctuating field, 
$\bfb(\bfX, \tau)$, whose evolution is determined by:
\begin{eqnarray}
\left(\frac{\partial}{\partial\tau} \,+\, SX_1\frac{\partial}{\partial X_2}\right)\bfb \;-\; Sb_1\ey &\;=\;& \bnabla\cross\left[\,\alpha \overline{\bfB}\,\right] \,+\,\bnabla\cross\left[\,\alpha\bfb-\overline{\alpha\bfb}\right]\,+\,\eta_T\bnabla^2\bfb\,, \nonumber\\[2ex]
\bnabla\cendot\bfb &\;=\;& 0\,,\quad \mbox{with initial condition}\quad\bfb(\bfX, 0)=\bf0\,.
\label{flucalpfluc}
\end{eqnarray}

\noindent
The physical assumption behind the choice of the initial condition $\bfb(\bfX, 0)=\bf0$ is that the homogeneous part of equation~(\ref{flucalpfluc}) has only decaying solutions. Then the origin of time can always be chosen such that the solution of the homogeneous part  have already decayed. Then the $\bfb(\bfX,\tau)$ which contributes to the EMF in equation~(\ref{meanEMF}) is the one whose source is $\bnabla\cross\left[\,\alpha \overline{\bfB}\,\right]$.

We now need to specify the statistics of the $\alpha$ fluctuations.
Shear flows possess a natural symmetry, called \emph{Galilean invariance}  in \citet{SS09a,SS09b}, which is related to measurements made by observers, whose velocity with respect to the lab frame $(\bfX, \tau)$ is equal to that of the background shear flow. All these \emph{comoving observers} may be labelled by $\bfxi=(\xi_1, \xi_2, \xi_3)\,$, the position of
the origin of the comoving observer at the initial time zero. The position of the origin of the comoving observer at time $\tau$ is given by
\beq
\bfX_c(\bfxi, \tau) \;=\; (\xi_1, \,\xi_2+S\tau \xi_1, \,\xi_3)\,.
\label{CO_org}
\eeq

\noindent
Galilean invariance (GI) is the property of the form--invariance of a
quantity that is transformed to the rest frame of any comoving observer, with $\bfxi \in \mathbb{R}^3\,$. For GI $\alpha$ fluctuations, an $n$--point correlator given in the lab frame must equal a correspondingly constructed $n$--point correlator in the frame of any comoving observer. Therefore, for GI $\alpha$ fluctuations, we must have:
\beq
\overline{\alpha(\bfX^{(1)}, \tau_1)\ldots \alpha(\bfX^{(n)}, \tau_n)} \;=\;
\overline{\alpha(\bfX^{(1)} +\bfX_c(\bfxi, \tau_1), \tau_1)\,\ldots\,
\alpha(\bfX^{(n)} +\bfX_c(\bfxi, \tau_n), \tau_n)}\,.
\label{GI_alp-alp}
\eeq

\noindent
It may be verified that, for GI $\alpha$ fluctuations, the dynamo equation~(\ref{totmf_ensavd_expl}), and equations~(\ref{meanalpfluc}) and (\ref{flucalpfluc}) for the large--scale and fluctuating fields, are Galilean Invariant.\footnote{We have argued that the GI 
$\alpha$ statistics of equation~(\ref{GI_alp-alp}) is the one that 
respects the natural symmetry of a linear shear flow. However, this 
symmetry can be broken, if the sources of the $\alpha$ fluctuations 
picked out a special frame. In this case, equation~(\ref{GI_alp-alp})
need not be true, and (\ref{totmf_ensavd_expl})---(\ref{flucalpfluc})
would no longer respect the symmetry of GI.}

\noindent
Equations~(\ref{totmf_ensavd_expl}), (\ref{meanalpfluc}) and (\ref{flucalpfluc}) do not depend explicitly on time, so if we also require 
that the $\alpha$ fluctuations have time--stationary statistics, then
we can expand the set of observers to include comoving observers whose origin of time may be shifted by an arbitrary constant amount. All comoving observers can be labelled by the coordinates, $(\bfxi, \tau_0)$, where $\tau_0$ is the time read by the clock of a comoving observer at position $\bfxi$ at time $0$.
Therefore $\alpha$ fluctuations that are GI and time--stationary must satisfy:
\begin{eqnarray}
&&\overline{\alpha(\bfX^{(1)}, \tau_1)\ldots \alpha(\bfX^{(n)}, \tau_n)}\nonumber\\
&&\qquad =\;\,\overline{\alpha(\bfX^{(1)} +\bfX_c(\bfxi, \tau_1+\tau_0), \tau_1+\tau_0) \ldots\alpha(\bfX^{(n)}+\bfX_c(\bfxi, \tau_n+\tau_0), \tau_n+\tau_0)}\,.
\label{StGI_alp-alp}
\end{eqnarray}
 
\noindent
It should be noted that the general requirements of Galilean--Invariance 
(and the additional specialization to time--stationarity), given in  
equations~(\ref{GI_alp-alp}) and (\ref{StGI_alp-alp}), will continue to 
hold, regardless of the details of an underlying more detailed theory
based on velocity correlators. 

Our model is now fully defined by the equations~(\ref{meanalpfluc}) and (\ref{flucalpfluc}) for the large--scale and fluctuating 
magnetic fields, $\overline{\bfB}(\bfX, \tau)$ and $\bfb(\bfX, \tau)$, together with either \emph{general Galilean--Invariant $\alpha$ statistics} (eqn.~\ref{GI_alp-alp}) or \emph{time--stationary Galilean--Invariant 
$\alpha$ statistics} (eqn.~\ref{StGI_alp-alp}). To derive a closed equation
for $\overline{\bfB}(\bfX, \tau)$, it is necessary to do the following: 
Solve equation~(\ref{flucalpfluc}) to get $\bfb(\bfX, \tau)$ as a
functional of $\alpha(\bfX, \tau)$ and $\overline{\bfB}(\bfX, \tau)$. Use this
in equation~(\ref{meanEMF}) to get the EMF $\bfemf(\bfX, \tau)$ as a
functional of $\overline{\bfB}(\bfX, \tau)$ and various $n$--point $\alpha$ correlators, which
are required to be either general GI (eqn.~\ref{GI_alp-alp}) or time--stationary GI (eqn.~\ref{StGI_alp-alp}).
The $\alpha$ correlators are specified quantities, so $\overline{\bfemf}$ can be
thought of as a functional only of $\overline{\bfB}$. Using this form
for $\overline{\bfemf}$ in equation (\ref{meanalpfluc}) will result in a closed integro--differential equation for $\overline{\bfB}\,$.

The above program is best realized in \emph{sheared coordinates} $(\bfx, t)$, which are defined in terms of the lab coordinates $(\bfX, \tau)$ as:
\beq
x_1\,=\,X_1\,;\qquad x_2\,=\,X_2-S\tau X_1\,;\qquad x_3\,=\,X_3\,;\qquad t\,=\,\tau\,.
\label{tr-sh}
\eeq

\noindent
The inverse transformation is:
\beq
X_1\,=\,x_1\,;\qquad X_2\,=\,x_2+St x_1\,;\qquad X_3\,=\,x_3\,;\qquad \tau\,=\,t\,.
\label{invtr-sh}
\eeq

\noindent
The $\bfx$ are the Lagrangian coordinates of fluid elements in the background
shear flow. Using equation~(\ref{CO_org}) for the definition of the
function $\bfX_c\,$, the shearing coordinate transformation and its inverse
can also be written as $\bfx = \bfX_c(\bfX, -\tau)\,$ and $\bfX = \bfX_c(\bfx, t)\,$. Since all comoving observers are equivalent, \emph{every comoving observer has corresponding sheared coordinates}. 

\noindent
We want to recast all the equations~(\ref{totmf_ensavd_expl})---(\ref{StGI_alp-alp}) in terms of new fields that are functions of $(\bfx, t)$. These are the meso--scale field $\bfH(\bfx, t)=\bfB(\bfX, \tau)\,$, the large--scale field $\overline{\bfH}(\bfx, t)=\overline{\bfB}(\bfX, \tau)$, the fluctuating field $\bfh(\bfx, t)=\bfb(\bfX, \tau)$, the 
fluctuating alpha $a(\bfx, t)=\alpha(\bfX, \tau)$,  and the large--scale EMF
$\overline{\bfE}(\bfx, t)=\overline{\bfemf}(\bfX, \tau)\,$.\footnote{The new vector fields are component--wise equal to the old vector fields, in that both are resolved along the fixed basis $(\ex, \ey, \ez)$. For instance, $H_1(\bfx, t) = B_1(\bfX, \tau)$ and so on.} Then equations~(\ref{totmf_ensavd_expl}), (\ref{meanalpfluc}) and (\ref{flucalpfluc}) give:
\beq
\frac{\partial \bfH}{\partial t} \;-\; SH_1\ey \;=\; \bnabla\cross\left[\,a \bfH\,\right]\,+\,\eta_T\bnabla^2\bfH\,,\qquad\qquad\bnabla\cendot\bfH = 0\,;
\label{DE-sh}
\eeq
\beq
\frac{\partial \overline{\bfH}}{\partial t} \;-\; S\overline{H_1}\ey \;=\;
\bnabla\cross\overline{\bfE} \,+\, \eta_T\bnabla^2\overline{\bfH}
\,,\qquad\qquad\bnabla\cendot\overline{\bfH} \;=\; 0\,,\qquad\qquad\overline{\bfE} \;=\;
\overline{a\bfh}\,;
\label{MFE-sh}
\eeq

\begin{eqnarray}
\frac{\partial \bfh}{\partial t} \;-\; Sh_1\ey &\;=\;&
\bnabla\cross\left[\,a\overline{\bfH}\,\right] \,+\,
\bnabla\cross\left[\,a\bfh-\overline{a\bfh}\right]\,+\,\eta_T\bnabla^2\bfh
\,,\nonumber\\[1ex]
\bnabla\cendot\bfh &\;=\;& 0\,,\quad \mbox{with initial condition}\quad\bfh(\bfx, 0)=\bf0\,;
\label{FFE-sh}\\[2ex]
\mbox{where} \qquad\bnabla &\;=\;& \frac{\partial}{\partial \bfx} \,-\, 
\ex St\frac{\partial}{\partial x_2}\qquad \mbox{is a time--dependent operator.}
\label{nabla-sh} 
\end{eqnarray}

\noindent
Equations~(\ref{DE-sh})---(\ref{FFE-sh}) are \emph{homogeneous} in $\bfx = (x_1, x_2, x_3)$, although explicitly dependent on $t$ through the operator
$\bnabla$ of equation~(\ref{nabla-sh}). We now rewrite the $n$--point 
correlators of equations~(\ref{GI_alp-alp}) and (\ref{StGI_alp-alp})
in the new variables. By definition, the
left hand sides of both equations are
$\;\overline{\alpha(\bfX^{(1)}, \tau_1)\ldots \alpha(\bfX^{(n)}, \tau_n)}\,=\,\overline{a(\bfx^{(1)}, t_1)\ldots a(\bfx^{(n)}, t_n)}\;$. 
The right hand sides can be 
worked out in a straightforward manner by applying the transformations of (\ref{tr-sh}) and (\ref{invtr-sh}), and the results stated simply:
\begin{itemize}
\item[(i)] 
A general Galilean--Invariant $n$--point correlator satisfies:
\beq
\overline{a(\bfx^{(1)}, t_1)\ldots a(\bfx^{(n)}, t_n)} \;=\;
\overline{a(\bfx^{(1)}+\bfxi, t_1)\ldots a(\bfx^{(n)}+\bfxi, t_n)}\,,
\label{GI_a-a}
\eeq

\noindent
for all $\bfxi\in\mathbb{R}^3\,$. Therefore, \emph{general GI implies 
spatial homogeneity of the correlators in sheared coordinates}. This 
is entirely consistent with the homogeneity of equations~(\ref{DE-sh})---(\ref{FFE-sh}) for the magnetic fields.

\item[(ii)]
If a Galilean--Invariant $n$--point correlator has the additional 
symmetry of being time--stationary, then:
\beq
\overline{a(\bfx^{(1)}, t_1)\ldots a(\bfx^{(n)}, t_n)} \;=\;
\overline{a(\bfx^{(1)}+\bfxi-S\tau_0 x^{(1)}_1\ey, t_1+\tau_0)\ldots
a(\bfx^{(n)}+\bfxi-S\tau_0 x^{(n)}_1\ey, t_n+\tau_0)}\,,
\label{StGI_a-a}
\eeq

\noindent
for all $\bfxi\in\mathbb{R}^3$ and $\tau_0\in\mathbb{R}\,$. Equations~(\ref{DE-sh})---(\ref{StGI_a-a}) complete the definition of our model in sheared coordinates.
\end{itemize}

Henceforth we develop the theory for \emph{time--stationary GI statistics}, 
for which the $n$--point correlators obey equation~(\ref{StGI_a-a}).
We first note that, for $\tau_0=0$, this reduces to (\ref{GI_a-a}): 
therefore every time--stationary GI correlator is also spatially homogeneous, when the base temporal points $(t_1, \ldots, t_n)$ are not shifted. But $\tau_0$ need not be zero; it can take any real value. When the base temporal points are allowed to be shifted by $\tau_0\neq 0$, the constraints on a time--stationary GI (instead of a general GI) $n$--point correlator are more severe. It must have the property of invariance under unequal displacements of the spatial base points $\left(\bfx^{(1)},\ldots,\bfx^{(n)}\right)$. This is a richer, 4--dimensional set of allowed variations, indexed by $(\bfxi, \tau_0)$, and hence is satisfied by only a subset of all the GI functions obeying equation~(\ref{GI_a-a}). This stronger constraint has entered the problem due to the additional requirement of time--stationarity. The restriction imposed is, indeed, strong enough to determine completely useful forms of the lowest order correlators. The $1$--point correlator $\overline{a(\bfx, t)} = 0$ by assumption. From equation~(\ref{StGI_a-a}), the time--stationary GI $2$--point correlator must satisfy:
\beq
\overline{a(\bfx, t)a(\bfx', t')} \;=\;
\overline{a(\bfx +\bfxi-S\tau_0 x_{1}\ey, t+\tau_0)
a(\bfx' +\bfxi-S\tau_0 x'_{1}\ey, t'+\tau_0)}\,,
\nonumber
\eeq

\noindent 
for all $\bfxi\in\mathbb{R}^3$ and $\tau_0\in\mathbb{R}\,$, so it is also true for the particular values $\bfxi = -\bfx' - St' x'_{1}\ey$ and 
$\tau_0 = -t'\,$. Then $\overline{a(\bfx, t)a(\bfx', t')} = 
\overline{a(\bfx-\bfx'+St'(x_1-x_1')\ey,\, t-t')a({\bf 0}, 0)\,}$, so that 
the general form of the $2$--point correlator is:
\beq
\overline{a(\bfx, t) a(\bfx', t')} \;=\; {\cal F}(\bfx-\bfx'+St'(x_1-x_1')\ey,\, t-t')\,,
\label{StGI-2pt}
\eeq

\noindent
where the \emph{$2$--point correlation function} ${\cal F}(\bfR, s)$ is a 
real function of one vector and one scalar argument. As expected from 
equation~(\ref{GI_a-a}), the correlation function is homogeneous, and depends on $\bfx$ and $\bfx'$ only through their difference $(\bfx - \bfx')$. However, it is not time stationary in the sheared coordinates.\footnote{By construction the $2$--point correlator is, indeed, time stationary in 
the lab frame $(\bfX, \tau)$ and all the associated comoving observers.
To see this, use (\ref{tr-sh}) in (\ref{StGI-2pt}): the left hand side is 
$\overline{\alpha(\bfX, \tau)\ldots \alpha(\bfX', \tau')}$ by definition. 
The right hand side is ${\cal F}\left(\bfX -\bfX' -S(\tau -\tau')X_1\ey\,, 
\tau -\tau'\right)$, so the correlation function is time--stationary
in the lab coordinates but no longer homogeneous, as expected.}
Henceforth we use the factored form, ${\cal F}(\bfR, t) = {\cal A}(\bfR)\,{\cal D}(t)\,$, to make contact with equation~(\ref{KMcorr}) used in the earlier discussion of the Kraichnan--Moffatt model without shear. Then 
\begin{eqnarray}
\qquad\qquad\overline{a(\bfx, t) a(\bfx', t')} &\;=\;& 2{\cal A}\!\left(\bfx-\bfx'+St'(x_1-x_1')\ey\right)\,{\cal D}(t-t')\,,\nonumber\\[1em]
\qquad\qquad2\int_0^{\infty}\,{\cal D}(t){\rm d}t &\;=\;& 1\,,\qquad\qquad
{\cal A}({\bf 0}) \;=\; \eta_{\alpha}\,>\,0\,.
\label{KMsh-2pt}
\end{eqnarray}
 
\noindent
We also define the  correlation time for the $\alpha$ fluctuations:
\beq
\tau_{\alpha} \;=\; 2\int_0^{\infty} \mathrm{d}t\;t\; {\cal D}(t)\,.
\label{corr-time}
\eeq

\noindent
The spatial correlation function is ${\cal A}(\bfR)$, where $\bfR = \bfx-\bfx'+St'(x_1-x_1')\ey$ is the relative separation vector of fluid elements
due to shear. In the absence of shear, $S=0$, the separation $\bfR = \bfx -
\bfx'$, and the right hand side of (\ref{KMsh-2pt}) is $2{\cal A}(\bfx - \bfx'){\cal D}(t - t')= 2{\cal A}(\bfX - \bfX'){\cal D}(\tau - \tau')\,$.
Hence, as expected, for zero shear, equation (\ref{KMsh-2pt}) reduces to equation~(\ref{KMcorr}), which applies to the Kraichnan--Moffatt model described earlier. For non zero shear the separation $\bfR$ is a time--dependent vector. Furthermore, for a general function ${\cal A}(\bfR)$ and $St' \neq 0$, it is evident that the $2$--point correlator will be an anisotropic and time--dependent function in $(\bfx - \bfx')$.

Having presented our model in detail, it is useful to review the principal assumptions and limitations. The model is intended to be a minimal extension of the KM model of \S~2.1 to include a linear shear, and we have followed suit in making the simplifying assumptions of locality and isotropy. By 
this we mean: in the  defining equation~(\ref{totmf_ensavd_expl}) for the meso--scale field, the relationship between the EMF and the meso--scale field is local and instantaneous, and the transport coefficients are isotropic. Specifically, EMF $\bfemf(\bfX, \tau)=\alpha(\bfX, \tau) \bfB-\eta_t \bnabla\cross \bfB$, where $\eta_t$ is a scalar constant, and $\alpha$ is a pseudo--scalar stochastic function with GI statistics. The $\alpha$ fluctuations have a 2--point correlator which is taken to be of the factored form in equation~(\ref{KMsh-2pt}). No restriction is placed on the forms of the functions $\bfA(\bfR)$ and ${\cal D}(t)$: they could be either dependent on shear or not; it is just that our model does not specify this. 

\section{Equation for the large--scale magnetic field}

In order to derive a closed equation for the large--scale magnetic field we exploit the homogeneity of the problem in the sheared coordinates $\bfx$, and work with Fourier variables $\bfk\in\mathbb{R}^3$ which are conjugate to them. Given any ${\cal T}(\bfx, t)$ --- which could be a (pseudo) scalar, vector or tensor function --- let $\widetilde{{\cal T}}(\bfk, t) = \int \mathrm{d}^3x \exp{(-\mathrm{i}\,\bfk\cendot\bfx)}\,{\cal T}(\bfx, t)\,$ be its Fourier transform. Thus we use $\left\{\widetilde{\bfH}(\bfk, t)\,, \widetilde{a}(\bfk, t)\,,\widetilde{\overline{\bfH}}(\bfk, t)\,,\widetilde{\bfh}(\bfk, t)\,,\widetilde{\overline{\bfE}}(\bfk, t)\right\}$ to denote the Fourier transforms of $\left\{\bfH(\bfx, t)\,, a(\bfx, t)\,,\overline{\bfH}(\bfx, t)\,, \bfh(\bfx, t)\,, \overline{\bfE}(\bfx, t)\right\}\,$. Similarly, we can also define the Fourier variables $\bfK\in\mathbb{R}^3$, which are conjugate to the fixed coordinates $\bfX\,$. Then, given any ${\cal T}(\bfX, \tau)$ --- which could be a (pseudo) scalar, vector or tensor function --- let $\widetilde{{\cal T}}(\bfK, \tau) = \int \mathrm{d}^3X \exp{(-\mathrm{i}\,\bfK\cendot\bfX)}\,{\cal T}(\bfX, \tau)\,$ be its Fourier transform. 

The first step is to solve for $\widetilde{\bfh}(\bfk, t)$ as a functional of the stochastic field $\widetilde{a}(\bfk, t)$ and the large--scale magnetic field $\widetilde{\overline{\bfH}}(\bfk, t)$. This is carried out below in the first--order smoothing approximation (FOSA), by dropping the terms $\bnabla\cross\left[\,a\bfh-\overline{a\bfh}\right]\,$  in equation~(\ref{FFE-sh}). We also drop $\,\eta_T\bnabla^2\bfh$, although this 
is not a necessary feature of FOSA. Hence:
\beq
\frac{\partial \bfh}{\partial t} \;-\; Sh_1\ey \;=\; \bnabla\cross\bfM\,,
\qquad\qquad\bnabla\cendot\bfh \;=\; 0\,,\qquad\qquad\bfh(\bfx, 0) \;=\; \bf0\,,
\label{FFE-fosa}
\eeq

\noindent
where $\bfM(\bfx, t) = a(\bfx, t)\overline{\bfH}(\bfx, t)$ is a 
stochastic source field, and $\bnabla$ is the timeÑ-dependent operator defined in (\ref{nabla-sh}). Fourier transforming, we get:
\begin{eqnarray}
\frac{\partial \widetilde{\bfh}}{\partial t} \;-\; S\widetilde{h_1}\ey
&\;=\;& \mathrm{i}\bfK\cross\widetilde{\bfM}\,,\qquad
\bfK\cendot\widetilde{\bfh} \;=\; 0\,,\qquad \widetilde{\bfh}(\bfk, 0) 
\;=\; \bf0\,,\nonumber\\[1em]
\bfK(\bfk, t) &\;=\;& 
\ex(k_1 - St\,k_2) \,+\, \ey k_2 \,+\, \ez k_3\,,\nonumber\\[1em]
\widetilde{\bfM}(\bfk, t) &\;=\;&  
\frac{1}{(2\pi)^3} \int \mathrm{d}^3k'\,{\widetilde{a}}^{\,*}(\bfk', t)\,
\widetilde{\overline{\bfH}}(\bfk+\bfk', t)\,.
\label{FFE-fou}
\end{eqnarray}

\noindent
Equation~(\ref{FFE-fou}) can be integrated directly componentÑwise, and the 
solution which satisfies both constraints, $\bfK\cendot\widetilde{\bfh} 
= 0$ and $\widetilde{\bfh}(\bfk, 0) = \bf0\,$, written as:
\beq
\widetilde{\bfh}(\bfk, t) \;=\; \int_0^t \mathrm{d}t'\,
\left[\mathrm{i}\bfK(\bfk, t')\cross\widetilde{\bfM}(\bfk, t')\right] \;+\;
\ey S\int_0^t \mathrm{d}t'\int_0^{t'} \mathrm{d}t''\,
\left[\mathrm{i}\bfK(\bfk, t'')\cross\widetilde{\bfM}(\bfk, t'')\right]_1
\eeq

\noindent
The double time integral can be reduced to single time integral by noting
that:
\beq
\int_0^t \mathrm{d}t'\int_0^{t'} \mathrm{d}t''\,f(t'') \;=\;
\int_0^t \mathrm{d}t' (t-t') f(t')\,.
\nonumber
\eeq

\noindent
Then the FOSA solution for the fluctuating magnetic field can be written as:
\beq
\widetilde{\bfh}(\bfk, t) \;=\; \int_0^t \mathrm{d}t'\,\left\{\,
\mathrm{i}\bfK(\bfk, t')\cross\widetilde{\bfM}(\bfk, t') \;+\; 
\ey S(t-t')\left[\mathrm{i}\bfK(\bfk, t')\cross\widetilde{\bfM}(\bfk, t')
\right]_1\,\right\}\,.
\label{Fhsoln}
\eeq

\noindent
Equation~(\ref{Fhsoln}) should be used to calculate the Fourier transform
of the EMF:
\begin{eqnarray}
\widetilde{\overline{\bfE}}(\bfk, t) &=&   \int \mathrm{d}^3x
\exp{(-\mathrm{i}\,\bfk\cendot\bfx)}\,\overline{\bfE}(\bfx, t)
=\int \mathrm{d}^3x
\exp{(-\mathrm{i}\,\bfk\cendot\bfx)}\,\overline{a(\bfx, t)\,\bfh(\bfx, t)}\nonumber \\[2ex]
&=& \frac{1}{(2\pi)^3} \int \mathrm{d}^3k_1\, \mathrm{d}^3k_2\,
\delta(\bfk_1+\bfk_2-\bfk)\;\overline{\widetilde{a}(\bfk_1, t)\,\widetilde{\bfh}(\bfk_2, t)}
\nonumber \\[2ex]
&=& \frac{1}{(2\pi)^3} \int \mathrm{d}^3k_1\, \mathrm{d}^3k_2\,
\delta(\bfk_1+\bfk_2-\bfk)\;\int_0^t \mathrm{d}t'\,\Biggl\{\, 
\left[\,\mathrm{i}\bfK(\bfk_2, t')\cross\overline{\widetilde{a}(\bfk_1, t)
\widetilde{\bfM}(\bfk_2, t')}\,\right]\nonumber\\[2ex]
&&\quad\quad\;+\;\ey S(t-t')\left[\,\mathrm{i}\bfK(\bfk_2, t')\cross
\overline{\widetilde{a}(\bfk_1, t)\widetilde{\bfM}(\bfk_2, t')}\,\right]_1
\,\Biggr\}\,,\label{EMFfou} 
\end{eqnarray}

\noindent
is given in terms of the quantity $\overline{\widetilde{a}(\bfk_1, t)\widetilde{\bfM}(\bfk_2, t')}\,$, which has to be calculated. 
Using equation~(\ref{FFE-fou}) for $\widetilde{\bfM}\,$,
\beq
\overline{\widetilde{a}(\bfk_1, t)\widetilde{\bfM}(\bfk_2, t')} \;=\;
\frac{1}{(2\pi)^3} \int \mathrm{d}^3k_3\,\overline{\widetilde{a}(\bfk_1, t)
\widetilde{a}^*(\bfk_3, t')}\;\widetilde{\overline{\bfH}}(\bfk_2+\bfk_3, t')
\label{am-eqn}
\eeq

\noindent 
is a convolution of the large--scale magnetic field and the Fourier--space $2$--point correlator. We now compute the latter using equation~(\ref{KMsh-2pt}) for time--stationary  Galilean--Invariant $\alpha$ fluctuations:
\begin{eqnarray}
&&\overline{\widetilde{a}(\bfk_1, t)\widetilde{a}^*(\bfk_3, t')} \;=\;
\int \mathrm{d}^3x_1 \mathrm{d}^3x_3 \exp{(-\mathrm{i}\,\bfk_1\cendot\bfx_1
+\mathrm{i}\,\bfk_3\cendot\bfx_3)}\;\overline{a(\bfx_1, t) a(\bfx_3, t')}\nonumber\\[2ex]
&&\;=\; 2{\cal D}(t-t')\int \mathrm{d}^3x_1 \mathrm{d}^3x_3 \exp{[-\mathrm{i}(\bfk_1\cendot\bfx_1
-\bfk_3\cendot\bfx_3)]}\,
{\cal A}\!\left(\bfx_1-\bfx_3+St'(x_{11}-x_{31})\ey\right)\,.
\nonumber
\end{eqnarray}

\noindent
Using new integration variables, $\bfr=\bfx_1-\bfx_3\,$ and 
$\bfr'=\frac{1}{2}(\bfx_1+\bfx_3)\,$, we get
\begin{eqnarray}
\overline{\widetilde{a}(\bfk_1, t)\widetilde{a}^*(\bfk_3, t')} &=&
2{\cal D}(t-t')\int \mathrm{d}^3r\, \mathrm{d}^3r' \exp{\left[\,-\mathrm{i}(\bfk_1-\bfk_3)
\cendot\bfr' - \frac{\mathrm{i}}{2}(\bfk_1+\bfk_3)\cendot\bfr\,\right]}
\; \times\nonumber\\[2ex]
&&\qquad\qquad\qquad\times\;{\cal A}\!\left(\bfr+St'r_1\ey\right)\nonumber\\[2ex]
&=& 2{\cal D}(t-t')(2\pi)^3\, \delta(\bfk_1-\bfk_3)\int \mathrm{d}^3r\,
\exp{(-\mathrm{i}\,\bfk_1\cendot\bfr)}\,{\cal A}\!\left(\bfr+St'r_1\ey\right)\,.
\nonumber
\end{eqnarray}

\noindent
Another change of the integration variable to $\bfR=\bfr+St'r_1\ey\,$
gives us a compact form for the $2$--point correlator:
\begin{eqnarray}
\overline{\widetilde{a}(\bfk_1, t)\widetilde{a}^*(\bfk_3, t')} &\;=\;&
2{\cal D}(t-t')(2\pi)^3\, \delta(\bfk_1-\bfk_3)\,\widetilde{{\cal A}}\left(\bfK(\bfk_1, t')\right)\,,\nonumber\\[1em]
\mbox{where}\quad 
\widetilde{{\cal A}}(\bfK) &\;=\;& \int \mathrm{d}^3R\,
\exp{(-\mathrm{i}\,\bfK\cendot\bfR)}\; {\cal A}(\bfR)\,.
\label{2ptfou}
\end{eqnarray}

\noindent
$\widetilde{{\cal A}}(\bfK)$ is the \mbox{complex spatial power spectrum} of $\alpha$ fluctuations, 
with $\widetilde{{\cal A}}(-\bfK) = \widetilde{{\cal A}}^*(\bfK)$ because
${\cal A}(\bfR)$ is a real function. The $\delta$--function in 
equation~(\ref{2ptfou}) implies that a shearing wave labelled by $\bfk_1$
is uncorrelated with any other shearing wave with $\bfk_1\neq \bfk_3\,$.
Also each shearing wave has a time--dependent complex power because
the argument of $\widetilde{{\cal A}}$ is a time--dependent wavevector, although the time correlation function ${\cal D}$ is stationary. \emph{Thus the $\alpha$
fluctuations may be thought of as a superposition of random and 
mutually uncorrelated shearing waves.} Using equation~(\ref{2ptfou}) 
in (\ref{am-eqn}), we get a simple expression:
\beq
\overline{\widetilde{a}(\bfk_1, t)\widetilde{\bfM}(\bfk_2, t')} \;=\;
2{\cal D}(t-t')\widetilde{{\cal A}}\left(\bfK(\bfk_1, t')\right)\,
\widetilde{\overline{\bfH}}(\bfk_1+\bfk_2, t')\,.
\label{am-fin}
\eeq

\noindent
When equation~(\ref{am-fin}) is substituted in (\ref{EMFfou}) we obtain a 
compact expression for the EMF:
\beq
\widetilde{\overline{\bfE}}(\bfk, t) \;=\; 2\int_0^t \mathrm{d}t'\,{\cal D}(t-t')\left\{\,\widetilde{\bfU}(\bfk, t')\cross\widetilde{\overline{\bfH}}(\bfk, t') \;+\; 
\ey S(t-t')\left[\widetilde{\bfU}(\bfk, t')\cross\widetilde{\overline{\bfH}}(\bfk, t')
\right]_1\,\right\}\,,
\label{EMF-U}
\eeq

\noindent
where 
\beq
\widetilde{\bfU}(\bfk, t') \;=\; 
\int\frac{\mathrm{d}^3k_1}{(2\pi)^3}\,\mathrm{i}\bfK(\bfk-\bfk_1, t')
\widetilde{{\cal A}}\left(\bfK(\bfk_1, t')\right)\,,
\eeq

\noindent 
is a  complex vector field that can be written
in terms of the quantities, $\eta_{\alpha}$ and $\bfV_{\!\!M}$,  
appearing in the KM theory without shear. Using the linearity of the 
function $\bfK\,$: i.e.  $\bfK(\bfk-\bfk', t') = \bfK(\bfk, t') - \bfK(\bfk', t')$, we see that: 
\begin{eqnarray}
\widetilde{\bfU}(\bfk, t') &\;=\;& 
\mathrm{i}\bfK(\bfk, t')\int\frac{\mathrm{d}^3k'}{(2\pi)^3}\,
\widetilde{{\cal A}}\left(\bfK(\bfk', t')\right) \;-\; 
\int\frac{\mathrm{d}^3k'}{(2\pi)^3}\,\mathrm{i}\bfK(\bfk', t')
\widetilde{{\cal A}}\left(\bfK(\bfk', t')\right)
\nonumber\\[1em]
&\;=\;& \mathrm{i}\bfK(\bfk, t')\int\frac{\mathrm{d}^3K'}{(2\pi)^3}\,
\widetilde{{\cal A}}\left(\bfK'\right) \;-\;
\int\frac{\mathrm{d}^3K'}{(2\pi)^3}\,\mathrm{i}\bfK'
\widetilde{{\cal A}}\left(\bfK'\right)
\nonumber\\[1em]
&\;=\;& \mathrm{i}\bfK(\bfk, t'){\cal A}({\bf 0}) \;-\; 
\frac{\partial{\cal A}(\bfxi)}{\partial\bfxi}{\Biggl {\vert}}_{\bfxi = {\bf 0}}
\;\;\;=\;\; \mathrm{i}\bfK(\bfk, t')\eta_{\alpha} \;+\; \bfV_{\!\!M}\,,
\label{Uform}
\end{eqnarray}

\noindent
where we have transformed to a new integration variable $\bfK' =\bfK(\bfk', t') = \left(k'_1 - St'k'_2\,, k'_2\,, k'_3\right)\,$ --- which has
unit Jacobian giving $\mathrm{d}^3k'=\mathrm{d}^3K'$ --- and used the definitions of $\eta_{\alpha}$ and $\bfV_{\!\!M}$ given in equations~(\ref{KMcorr}) and (\ref{KMmfe}). 

Substituting (\ref{Uform}) for $\widetilde{\bfU}(\bfk, t')$ in (\ref{EMF-U}), we obtain an explicit expression for the EMF: 
\begin{eqnarray}
\widetilde{\overline{\bfE}}(\bfk, t) &\;=\;& 2\eta_{\alpha}\int_0^t \mathrm{d}t'\,{\cal D}(t-t')\left\{\,\mathrm{i}\bfK(\bfk, t')\cross\widetilde{\overline{\bfH}}(\bfk, t') \;+\; \ey S(t-t')\left[\mathrm{i}\bfK(\bfk, t')\cross\widetilde{\overline{\bfH}}(\bfk, t')
\right]_1\,\right\}\,,\nonumber\\[2ex]
&&\;+\quad
2\int_0^t \mathrm{d}t'\,{\cal D}(t-t')\left\{\,\bfV_{\!\!M}\cross\widetilde{\overline{\bfH}}(\bfk, t') \;+\; \ey S(t-t')\left[\bfV_{\!\!M}\cross\widetilde{\overline{\bfH}}(\bfk, t')
\right]_1\,\right\}\,.
\label{EMF-etav}
\end{eqnarray}

\noindent
Fourier transforming equation~(\ref{MFE-sh}), we the equation governing 
the large--scale field is:
\beq
\frac{\partial \widetilde{\overline{\bfH}}}{\partial t} \;-\;
S\widetilde{\overline{H}_1}\ey \;=\; \mathrm{i}\bfK(\bfk, t)\cross\widetilde{\overline{\bfE}} \;-\; \eta_T K^2(\bfk, t)\,\widetilde{\overline{\bfH}}\,,\qquad\qquad \bfK(\bfk, t)\cendot \widetilde{\overline{\bfH}} \;=\; 0\,.
\label{MFE-fou}
\eeq

\noindent
\emph{Equation~(\ref{MFE-fou}), together with equation~(\ref{EMF-etav}) gives a closed, linear integro--differential equation for the large--scale magnetic field, $\widetilde{\overline{\bfH}}(\bfk, t)$, which is the principal general result of this paper. This mean--field equation depends on the three parameters $\left(\eta_{\alpha}\,, \bfV_{\!\!M}\,, S\right)\,$, as well as the time--correlation function ${\cal D}(t)\,$. It should be noted 
that the theory is likely to overestimate the growth rates of high wavenumber modes because the dissipative term, $\eta_T\bnabla^2\bfh\,$, was dropped in the FOSA equation~(\ref{FFE-fosa})}.

\section{White--noise $\alpha$ fluctuations}

For white--noise $\alpha$ fluctuations the normalized correlation function
is ${\cal D}_{\!{\rm WN}}(t) = \delta(t)\,$, the Dirac delta--function.
From equation~(\ref{corr-time}), we have $\tau_{\alpha} = 0\,$, which agrees with the notion of no memory (to explore memory effects it is necessary to consider ${\cal D}(t) \neq \delta(t)\,$ which, in general, will have $\tau_\alpha \neq 0$, and this is taken up later). In the rest of this section we study dynamo action due to white--noise $\alpha$ fluctuations.
From equation~(\ref{EMF-etav}) the EMF for white--noise is:
\beq
\widetilde{\overline{\bfE}}_{\rm WN}(\bfk, t) 
\;=\; \left[\,\mathrm{i}\bfK(\bfk, t)\eta_{\alpha} \;+\; \bfV_{\!\!M}\,\right]\cross\widetilde{\overline{\bfH}}(\bfk, t)\,,
\label{EMF-wn-fou}
\eeq

\noindent
which depends on the large--scale field at the present time only, because
white--noise carries no memory of the past. Substituting (\ref{EMF-wn-fou}) in (\ref{MFE-fou}), we get the following partial differential equation for the large--scale magnetic field:
\beq
\frac{\partial \widetilde{\overline{\bfH}}}{\partial t} \;+\;
\left[\,\eta_K K^2 \,+\, \mathrm{i}\, \bfK\cendot\bfV_{\!\!M}\,\right]\,
\widetilde{\overline{\bfH}}
\;=\; S\widetilde{\overline{H}_1}\ey\,,
\qquad\qquad \bfK(\bfk, t)\cendot\widetilde{\overline{\bfH}} \;=\; 0\,.
\label{MFE-wn}
\eeq

\noindent
We can solve this by defining a new field $\widetilde{\overline{\bfF}}(\bfk, t)$ through:
\beq
\widetilde{\overline{\bfH}}(\bfk, t) \;=\;
\widetilde{{\cal G}}(\bfk, t)\,\widetilde{\overline{\bfF}}(\bfk, t)\,,
\label{HG}
\eeq

\noindent
where
\begin{eqnarray}
\widetilde{{\cal G}}(\bfk, t) &\;=\;& \exp{\left\{\,-\int_0^t\,\mathrm{d}t' 
\left[\,\eta_K K^2(\bfk, t')\,+\,\mathrm{i}\, \bfV_{\!\!M}\cendot\bfK(\bfk, t')\,\right]\,\right\}}\nonumber\\[1em]
&\;=\;& \exp{\Bigl\{-\eta_K\left[\,k^2t - Sk_1k_2t^2 + (S^2/3)k_2^2t^3\,\right] \,-\, \mathrm{i}\left[\,\left(\bfV_{\!\!M}\cendot\bfk\right) t - (S/2) V_{\!M1}k_2 t^2\,\right]\Bigr\}}\,,
\nonumber\\
&&\label{greenfn}
\end{eqnarray}

\noindent
is a shear--diffusive Green's function in which the Kraichnan diffusivity
$\eta_K$ contributes to the amplitude, and the Moffatt drift velocity
$\bfV_{\!\!M}$ contributes only to the phase. Using (\ref{HG}) and (\ref{greenfn}) in (\ref{MFE-wn}), we see that $\widetilde{\overline{\bfF}}(\bfk, t)$ must obey
\beq
\frac{\partial \widetilde{\overline{\bfF}}}{\partial t} \;=\;
S\widetilde{\overline{F}}_1\ey\,,
\qquad\qquad \bfK(\bfk, t)\cendot\widetilde{\overline{\bfF}}(\bfk, t) \;=\; 0\,.
\label{Feq}
\eeq

\noindent
The solution is given by $\widetilde{\overline{\bfF}}(\bfk, t)=\widetilde{\overline{\bfF}}(\bfk, 0)+\ey St\,\widetilde{\overline{F}}_1(\bfk, 0)\,$, 
where $\widetilde{\overline{\bfF}}(\bfk, 0)$ is any vector which satisfies 
$\bfk\cendot\widetilde{\overline{\bfF}}(\bfk, 0) = 0\,$. From (\ref{HG}) we have $\widetilde{\overline{\bfH}}(\bfk, 0)=\widetilde{\overline{\bfF}}(\bfk, 0)$, so the solution to (\ref{MFE-wn}) may be written as:
\beq
\widetilde{\overline{\bfH}}(\bfk, t) \;=\;
\widetilde{{\cal G}}(\bfk, t)\,\left[\,
\widetilde{\overline{\bfH}}(\bfk, 0)+\ey St\,\widetilde{\overline{H}}_1(\bfk, 0)\,\right]\,,\qquad\qquad\bfk\cendot
\widetilde{\overline{\bfH}}(\bfk, 0) \;=\; 0\,.
\label{Hsoln-wn}
\eeq

\noindent
Whereas (\ref{Hsoln-wn}) is a complete and compact form of the solution to the problem of white--noise $\alpha$ fluctuations, it is also useful to view the problem in real space, instead of Fourier space. We begin with the expression for the EMF in real space. Using (\ref{EMF-wn-fou}), 
and the fact that $\mathrm{d}^3k=\mathrm{d}^3K\,$, $\,\bfk\cendot\bfx=\bfK\cendot\bfX\,$ and $\widetilde{\overline{\bfH}}(\bfk, t)=\widetilde{\overline{\bfB}}(\bfK, \tau)\,$, we have:
\begin{eqnarray}
 \overline{\bfemf}_{\rm WN}(\bfX, \tau) &\;=\;& 
 \overline{\bfE}_{\rm WN}(\bfx, t) \;=\;
 \int \frac{\mathrm{d}^3k}{(2\pi)^3}
 \exp{(\mathrm{i}\bfk\cendot\bfx)}\,
 \widetilde{\overline{\bfE}}_{\rm WN}(\bfk, t) \nonumber\\[1em]
 &\;=\;& \int \frac{\mathrm{d}^3K}{(2\pi)^3}
 \exp{(\mathrm{i}\bfK\cendot\bfX)}\,
 \left[\,\mathrm{i}\bfK \eta_{\alpha} \;+\; \bfV_{\!\!M}\,\right] 
 \cross\widetilde{\overline{\bfB}}(\bfK, \tau) \nonumber\\[1em]
 &\;=\;& \eta_{\alpha}\bnabla\cross\overline{\bfB} \;+\;
 \bfV_{\!\!M}\cross\overline{\bfB}\,.
 \label{EMF-wn}
\end{eqnarray}

\noindent
\emph{Thus the EMF for white--noise alpha fluctuations with shear is identical in form to the EMF for the Kraichnan--Moffatt case without shear. The physical reason why shear does not explicitly contribute to 
$\,\overline{\bfemf}_{{\rm WN}}$ is that white--noise has zero correlation time, giving shear no time to act. This is true for any finite value of the rate of shear parameter.}

Substituting (\ref{EMF-wn}) in (\ref{meanalpfluc}), we get the
following partial differential equation for the large--scale
magnetic field:

\beq
\left(\frac{\partial}{\partial\tau} \,+\, SX_1\frac{\partial}{\partial X_2}\right)
\overline{\bfB} - S\overline{B_1}\ey \;=\;
\bnabla\cross\left(\bfV_{\!\!M}\cross\overline{\bfB}\right) \,+\,
\eta_K\,\bnabla^2\overline{\bfB}\,,
\qquad\qquad \bnabla\cendot\overline{\bfB} \;=\; 0\,.
\label{meanalpfluc-wn}
\eeq

\noindent
The general solution of (\ref{meanalpfluc-wn}) can be written down, using the Fourier space solution (\ref{Hsoln-wn}) derived earlier:
\begin{eqnarray}
\overline{\bfB}(\bfX, \tau) &\;=\;& \overline{\bfH}(\bfx, t)\;=\; \int\frac{\mathrm{d}^3k}{(2\pi)^3}\exp{\left(\mathrm{i}\bfk\cendot\bfx\right)}\,\widetilde{\overline{\bfH}}(\bfk, t)
\nonumber\\[1em]
&\;=\;& \int \frac{\mathrm{d}^3k}{(2\pi)^3}
\exp{\left[\,\mathrm{i}\bfK(\bfk, \tau)\cendot\bfX\right]}\,
\widetilde{{\cal G}}(\bfk, \tau)\left[\,
\widetilde{\overline{\bfH}}(\bfk, 0)+\ey S\tau\,\widetilde{\overline{H}}_1(\bfk, 0)
\,\right]\,.\nonumber\\
&&\label{Bwn-slon}
\end{eqnarray}

\noindent
This is a superposition of shearing waves, whose growth or decay is controlled by the Green's function $\widetilde{{\cal G}}(\bfk, \tau)\,$.
We note some general properties:
\begin{itemize}
\item[1.] In the limit of no shear, $S=0\,$,  equation~(\ref{meanalpfluc-wn}) reduces to (\ref{KMmfe}). \emph{Thus the Kraichnan--Moffatt problem corresponds to the case of white--noise} ($\tau_\alpha = 0$) \emph{and no shear} ($S$).

\item[2.] {\bf Weak $\alpha$ fluctuations} have $\eta_{\alpha} < \eta_T\,$ 
so that $\eta_K > 0\,$. Waves of all wavenumbers $\bfk$ eventually decay. 
``Axisymmetric'' waves, whose shear--wise wavenumber $k_2 = 0\,$, decay 
as $\exp{\left[-\eta_Kk^2t\right]}\,$.\footnote{We use the 
term ``axisymmetric'' in the sense in which it is often used in local
treatments of disc--dynamos, where it is conventional to take 
$(\ex,\ey,\ez)$ to be unit vectors along the radial, azimuthal and vertical directions.} The asymptotic decay of non--axisymmetric waves ($k_2 \neq 0$) is more rapid, like $\exp{\left[-\eta_K(S^2/3)k_2^2t^3\right]}\,$.

\item[3.] {\bf Strong $\alpha$ fluctuations} have $\eta_{\alpha} > \eta_T\,$ so that $\eta_K < 0\,$. Waves of all wavenumbers $\bfk$ eventually grow.
Axisymmetric waves grow like $\exp{\left[\vert\eta_K\vert k^2t\right]}\,$,
due to negative diffusion. Non--axisymmetric waves grow even more rapidly, like $\exp{\left[\vert\eta_K\vert(S^2/3)k_2^2t^3\right]}\,$, where shear acts in conjunction with the negative diffusion due to $\alpha$ fluctuations.

\item[4.] The Moffatt drift velocity $\bfV_{\!\!M}$ contributes only to
the phase and does not determine the growth or decay of
the large--scale magnetic field.
\end{itemize}

\emph{Therefore, the necessary condition for dynamo action for white--noise $\alpha$ fluctuations is that they must be strong, i.e. $\eta_K < 0\,$. This condition is identical to the Kraichnan--Moffatt case of no shear. To explore the effects of the Moffatt drift $\bfV_{\!\!M}$ and Shear $S$
on the growth rate, it is necessary to consider $\tau_\alpha\neq 0$, 
thereby allowing for memory effects}. 

\section{Axisymmetric large--scale magnetic fields when
$\tau_{\alpha}$ is small}

We now ask: \emph{Is it possible that, when $\tau_{\alpha} \neq 0$, there is dynamo action even for weak $\alpha$ fluctuations, i.e. when 
$\eta_K > 0\,$}? To answer this question in generality, it is necessary to deal with the linear integro--differential equation defined by equations~(\ref{EMF-etav}) and (\ref{MFE-fou}); such an analysis is beyond the scope of this paper. In this section we derive the partial differential equation governing the evolution of large--scale magnetic field which evolves over
times much larger than $\tau_\alpha$. Then we answer the above question by considering the dynamo action of the memory effect due to non zero $\tau_\alpha$ and $\eta_\alpha$, but without either Moffatt drift or shear.

\subsection{Derivation of the governing equation}

We begin by noting that the normalized time correlation function, ${\cal D}(t)$, has a singular limit:  i.e. $\lim\limits_{\tau_{\alpha}\to 0}{\cal D}(t)={\cal D}_{\rm WN}(t)=\delta(t)$. The limit of small but non zero $\tau_{\alpha}$ corresponds to looking at functions ${\cal D}(t)$ that are, in some sense, close to the Dirac delta--function. However we are saved from the task of considering variations in function space, because the EMF depends on ${\cal D}(t)$ only through a time integral: 
\beq
\widetilde{\overline{\bfE}}(\bfk, t) \;=\; 2\int_0^t \mathrm{d}s\,{\cal D}(s)\left\{\,\widetilde{\bfU}(\bfk, t-s)\cross\widetilde{\overline{\bfH}}(\bfk, t-s) \;+\; 
\ey Ss\left[\widetilde{\bfU}(\bfk, t-s)\cross\widetilde{\overline{\bfH}}(\bfk, t-s)
\right]_1\,\right\}\,,
\label{EMF-U2}
\eeq

\noindent
is the same as equation~(\ref{EMF-U}), rewritten by changing the
integration variable from $t'$ to $s=t-t'\,$. The complex velocity field
$\widetilde{\bfU} \,=\, \mathrm{i}\bfK(\bfk, t')\eta_{\alpha} + \bfV_{\!\!M}$ is as defined in (\ref{Uform}). Since the limit $\lim\limits_{\tau_{\alpha}\to 0} \widetilde{\overline{\bfE}}(\bfk, t) = \widetilde{\overline{\bfE}}_{\rm WN}(\bfk, t) = \widetilde{\bfU}(\bfk, t)\cross\widetilde{\overline{\bfH}}(\bfk, t)$ is evidently non singular, we make the \emph{ansatz} that, for small $\tau_\alpha$, the EMF can be expanded in a power series in 
$\tau_{\alpha}$:
\beq
\widetilde{\overline{\bfE}}(\bfk, t) \;=\; \widetilde{\overline{\bfE}}_{\rm WN}(\bfk, t)
\,+\, \widetilde{\overline{\bfE}}^{(1)}(\bfk, t) \,+\,
\widetilde{\overline{\bfE}}^{(2)}(\bfk, t) \,+\, \,\dots
\label{Eexp}
\eeq

\noindent
where $\widetilde{\overline{\bfE}}_{\rm WN}(\bfk, t) \sim {\cal O}(1)$
and $\widetilde{\overline{\bfE}}^{(n)}(\bfk, t) \sim {\cal O}(\tau_{\alpha}^n)$ for $n \geq 1$. Below we verify this ansatz up to $n= 1$, for slowly varying magnetic fields.
  
We want to work out $\widetilde{\overline{\bfE}}(\bfk, t)$ to first order in $\tau_{\alpha}\,$, for $t\gg\tau_{\alpha}\,$. Since ${\cal D}(s)$ is strongly peaked for times $s\leq \tau_{\alpha}$ and becomes very small for larger 
$s$, most of the contribution to the integral in (\ref{EMF-U2}) comes only
from short times $0\leq s < \tau_{\alpha}$. Hence in (\ref{EMF-U2}) we can
(i) set the upper limit of the $s$--integral to $+\infty\,$; (ii) keep the terms inside the $\{\;\}$ in the integrand up to only first order in $s$.
We also note that, for $k_2\neq 0$, the magnitude of $\bfK(\bfk, t)=\ex(k_1 - St\,k_2) + \ey k_2 + \ez k_3$ increases without bound, leading to rapid time variations of $\widetilde{\overline{\bfH}}(\bfk, t)$. Henceforth we restrict attention to 
axisymmetric modes for which $k_2=0\,$. This implies that $\bfK(\bfk, t-s)=\bfK(\bfk, t)=\bfk=(k_1, 0, k_3)\,$, and $\widetilde{\bfU}(\bfk, t-s)=\widetilde{\bfU}(\bfk)=\eta_{\alpha}\mathrm{i}\,\bfk+\bfV_{\!\!M}$. We also write:
\beq
\widetilde{\overline{\bfH}}(\bfk, t-s) \;=\; \widetilde{\overline{\bfH}}(\bfk, t)
\,-\,s \frac{\partial \widetilde{\overline{\bfH}}(\bfk, t)}{\partial t} \,+\, \ldots\,.
\label{Taylor-H}
\eeq

\noindent
where it is assumed that $\bigl{\vert} \widetilde{\overline{\bfH}} \bigr{\vert} \,\gg\, s\bigl{\vert}\partial \widetilde{\overline{\bfH}}/\partial t\bigr{\vert} \,\gg\; s^2\bigl{\vert}\partial^2 \widetilde{\overline{\bfH}}/\partial t^2\bigr{\vert}\,,\, s^3\bigl{\vert}\partial^3 \widetilde{\overline{\bfH}}/\partial t^3\bigr{\vert}\,\mbox{etc}\,$. Then 
\beq
\{\;\}\;\mbox{of eqn.~(\ref{EMF-U2})} \;=\;  
\widetilde{\bfU}(\bfk)\cross\widetilde{\overline{\bfH}}(\bfk, t) \;+\; 
s\left\{-\widetilde{\bfU}\cross\frac{\partial \widetilde{\overline{\bfH}}}{\partial t}
\,+\,\ey S\left[\widetilde{\bfU}\cross\widetilde{\overline{\bfH}}\right]_1
\right\} \,+\, {\cal O}(s^2)\,.
\eeq

\noindent
The integral over $s$ can now be performed. Using the properties of 
${\cal D}(\tau)$, given in (\ref{KMcorr}) and (\ref{corr-time}), we get:
\beq
\widetilde{\overline{\bfE}}(\bfk, t) \;=\; \widetilde{\overline{\bfE}}_{\rm WN}(\bfk, t)
\,+\, \tau_{\alpha}
\left\{-\widetilde{\bfU}\cross\frac{\partial \widetilde{\overline{\bfH}}}{\partial t}
\,+\,\ey S\left[\widetilde{\bfU}\cross\widetilde{\overline{\bfH}}\right]_1
\right\} \,+\, {\cal O}(\tau_{\alpha}^2)\,.
\label{EMF-taualp1}
\eeq

\noindent
In order to get an expression for the EMF accurate to ${\cal O}(\tau_{\alpha})$, we need $\,\partial \widetilde{\overline{\bfH}}/\partial t\,$ only to ${\cal O}(\tau_{\alpha}^0) = {\cal O}(1)\,$. Using 
(\ref{EMF-taualp1}) in (\ref{MFE-fou}) we have:
\beq
\frac{\partial \widetilde{\overline{\bfH}}}{\partial t}\Biggr{\vert}_{{\cal O}(1)} \;=\;
S\widetilde{\overline{H}}_1\ey \;+\;
\mathrm{i}\bfk\cross\widetilde{\overline{\bfE}}_{\rm WN}(\bfk, t) \;-\;
\eta_Tk^2
\widetilde{\overline{\bfH}} 
\;=\; S\widetilde{\overline{H}_1}\ey \;-\;
\left[\,\mathrm{i}\bfk\cendot\widetilde{\bfU} \,+\, \eta_Tk^2\right]
\widetilde{\overline{\bfH}} \,.
\label{Hdot-0thord}
\eeq

\noindent
When this is used in equation~(\ref{EMF-taualp1}), we obtain an expression 
for the EMF,
\beq
\widetilde{\overline{\bfE}}(\bfk, t) \;=\;
\left[\,1 \,+\, \eta_T k^2 \tau_{\alpha} \,+\, \mathrm{i}\,(\bfk\cendot\widetilde{\bfU})\tau_{\alpha}\,\right]
\widetilde{\bfU}\cross\widetilde{\overline{\bfH}}
\;+\; S\tau_{\alpha}\left[\,\widetilde{\overline{H}}_1\ey\cross\widetilde{\bfU} \,+\, \left(\widetilde{\bfU}\cross\widetilde{\overline{\bfH}}\right)_1\!\ey\,\right]\,,
\label{EMF-taualp2}
\eeq

\noindent
accurate to ${\cal O}(\tau_{\alpha})$, which verifies the ansatz of equation~(\ref{Eexp}) up to $n = 1$, as claimed. Equation~(\ref{EMF-taualp2}) for the EMF is valid only when the large--scale magnetic field is slowly varying; to lowest order this condition can be stated as $\,\bigl{\vert} \widetilde{\overline{\bfH}} \bigr{\vert} \gg \tau_{\alpha}\bigl{\vert}\partial \widetilde{\overline{\bfH}}/\partial t \bigr{\vert}\,$. Using equation~(\ref{Hdot-0thord}) for $\,\partial \widetilde{\overline{\bfH}}/\partial t \,$, we see that the sufficient condition for equation~(\ref{EMF-taualp2}) to be valid is that the following three 
dimensionless quantities be small:
\beq
\vert S\tau_{\alpha}\vert \,\ll\, 1\,,\qquad\qquad
\vert \eta_K k^2 \tau_{\alpha}\vert \,\ll\, 1\,,\qquad\qquad
\vert k V_M \tau_{\alpha}\vert \,\ll\,1\,.
\label{cond}
\eeq

Substituting (\ref{EMF-taualp2}) in (\ref{MFE-fou}) we obtain: 
\begin{eqnarray}
\frac{\partial \widetilde{\overline{\bfH}}}{\partial t} &\;=\;&
\left[\,S\widetilde{\overline{H}}_1\ey \,-\, \eta_K k^2 \widetilde{\overline{\bfH}} \,-\,\mathrm{i}\,(\bfk\cendot\bfV_{\!\!M}) \widetilde{\overline{\bfH}}\,\right]
\left[\,1 \,+\, \mathrm{i}\,(\bfk\cendot\bfV_{\!\!M})\tau_{\alpha} \,-\,
\eta_{\alpha} k^2\tau_{\alpha}\,\right] 
\nonumber\\[1em] 
&&\qquad +\;\;S\tau_{\alpha} \left[\,V_{M2} \widetilde{\overline{H}}_3 \,-\,
V_{M3} \widetilde{\overline{H}}_2 \,-\,
\mathrm{i}\,\eta_{\alpha} k_3 \widetilde{\overline{H}}_2\,\right]
\left[\,-\mathrm{i}\,k_3\ex \,+\, \mathrm{i}\,k_1 \ez\,\right]\,,
\nonumber\\[1em]
&&\mbox{with}\quad \bfk \;=\; (k_1\,, 0\,, k_3)\,,\qquad \mbox{and}\qquad
k_1\widetilde{\overline{H}}_1 \,+\, k_3\widetilde{\overline{H}}_3 \;=\; 0\,.
\label{MFE-k2-0}
\end{eqnarray}

\noindent
\emph{Equation~(\ref{MFE-k2-0}) is the linear partial differential equation that determines the evolution of an axisymmetric, large--scale magnetic field which evolves over times that are much larger than $\tau_\alpha\,$.
The three drivers of this evolution are (i) the diffusivities $\eta_\alpha$ and $\eta_K$ ; (ii) the Moffatt drift $\bfV_{\!\!M}$; (iii) shear $S\,$. These parameters must satisfy the three conditions given in (\ref{cond}), which also involves the wavenumber $k\,$. Only the condition on shear is independent of the wavenumber $k\,$, and it implies the following: whereas the full equations~(\ref{MFE-fou} and \ref{EMF-etav})  are non--perturbative in both $S$ and $\tau_\alpha$, equation~(\ref{MFE-k2-0}) above is valid only when $\vert S\tau_{\alpha}\vert \,\ll\, 1\,$}.

It is readily verified that the $\tau_\alpha = 0$ limit of equation~(\ref{MFE-k2-0}) is identical to the $k_2=0$ case of equation~(\ref{MFE-wn}), which describes axisymmetric white--noise $\alpha$ fluctuations. Hence the 
solution can be obtained by setting $k_2=0$ in equations~(\ref{Hsoln-wn}) and (\ref{greenfn}). Here it is seen that shear contributes to the linear--in--time stretching of the field lines along the shear--wise ($\ey$) direction, and the Moffatt drift causes the field to oscillate sinusoidally at angular frequency $\omega = \bfk\cendot\bfV_{\!\!M}\,$. It is only the Kraichnan diffusivity that contributes to the growth rate, $\gamma = -\eta_K k^2\,$. Hence the necessary condition for dynamo action is that $\eta_K = \eta_T - \eta_\alpha < 0$, just like in the more general non--axisymmetric white--noise case treated in \S~4.  

\subsection{The Kraichnan problem with non zero $\tau_\alpha$}

\begin{figure}
\begin{center}
\includegraphics[width=\columnwidth]{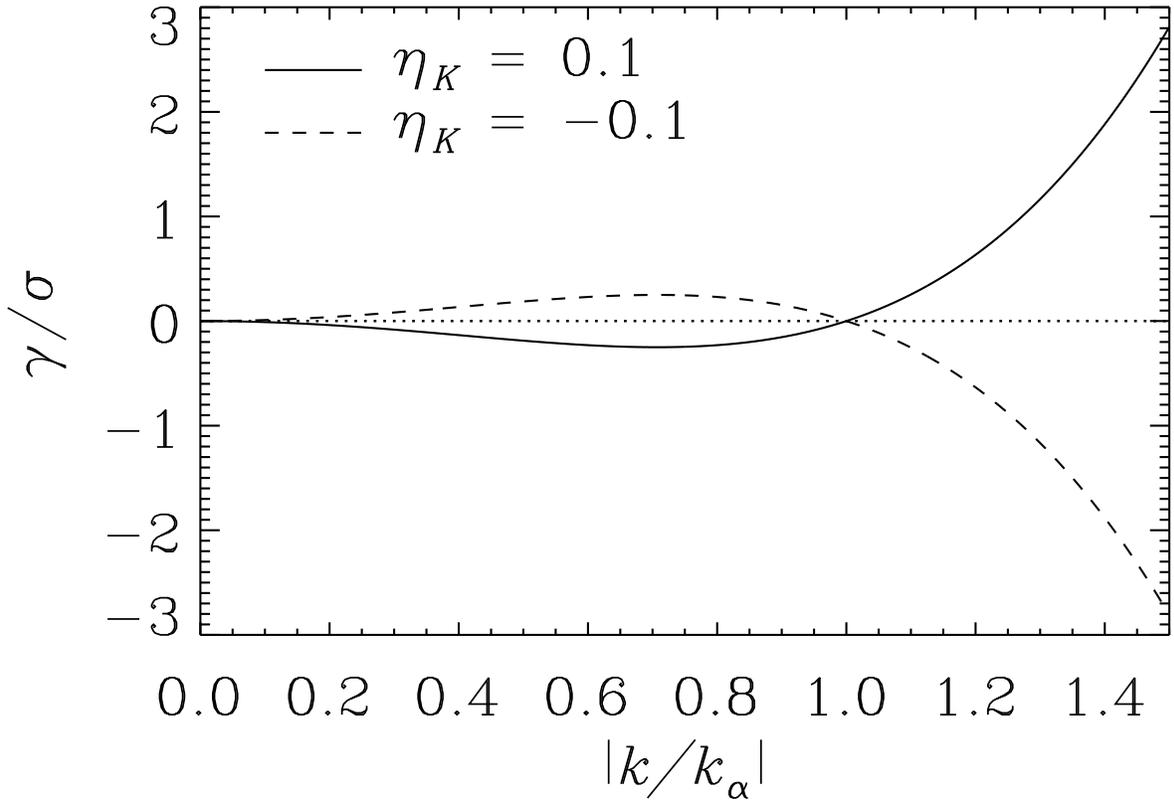}
\end{center}
\caption[]{Growth rate $\gamma/\sigma$ as a function of
$\vert k/k_{\alpha}\vert\,$, when $S=0$ and $\bfV_{\!\!M}= {\bf 0}\,$.
Weak ($\eta_K>0$) and strong ($\eta_K<0$)
$\alpha$ fluctuations are shown by bold and dashed curves, respectively.}
\label{Fig1}
\end{figure}

We want to understand the combined effect of the $\alpha$ fluctuations 
when $\bfV_{\!\!M}= {\bf 0}\,$ and $S=0$, but $\eta_\alpha > 0\,$ and $\tau_\alpha > 0\,$. Then there is a \emph{new length scale} in the problem,
whose corresponding wavenumber can be defined as:
\beq
k_\alpha \;=\; \left(\eta_\alpha\tau_\alpha\right)^{-1/2} \,\;>\;\, 0\,.
\label{kalpha}
\eeq

\noindent
When $S=0$ and $\bfV_{\!\!M}= {\bf 0}\,$, two of the three conditions in (\ref{cond}) are met trivially, and the third one implies that 
$\vert k\vert$ must be small enough such that $\vert\eta_K k^2\tau_\alpha\vert \ll 1\,$. However, $\vert k\vert$ can be larger or smaller than $k_\alpha\,$. We consider $\vert k\vert > k_\alpha$ to be \emph{high wavenumbers}, and $\vert k\vert < k_\alpha$ to be \emph{low wavenumbers}. 
Setting $S=0$ and $\bfV_{\!\!M}= {\bf 0}$ in equation~(\ref{MFE-k2-0}), we get:

\beq
\frac{\partial \widetilde{\overline{\bfH}}}{\partial t} \;=\;
\eta_K k^2\left[\left(\frac{k}{k_\alpha}\right)^2 \,-\, 1\right]\,
\widetilde{\overline{\bfH}}\,,\qquad
\mbox{with}\quad \bfk \;=\; (k_1\,, 0\,, k_3)\quad \mbox{and}\quad
k_1\widetilde{\overline{H}}_1 \,+\, k_3\widetilde{\overline{H}}_3 \;=\; 0\,.
\label{krcase}
\eeq

\noindent
This is the original Kraichnan problem for axisymmetric modes, modified by
a new factor given in $[\;]$. The solutions are of the exponential form, 
$\widetilde{\overline{\bfH}}(\bfk, t) \,=\, \widetilde{\overline{\bfH}}_0(\bfk)\exp{(\gamma t)}\,$, where $\bfk\cendot\widetilde{\overline{\bfH}}_0(\bfk) = 0\,$. Substituting this in equation~(\ref{krcase}), we get the 
growth rate as:
\beq
\gamma \;=\; \left(\eta_K k_\alpha^2\right)
\left(\frac{k}{k_\alpha}\right)^2
\left[\left(\frac{k}{k_\alpha}\right)^2 \,-\, 1\right]\,,\qquad\mbox{when}\qquad \vert\eta_K k^2\tau_\alpha\vert \ll 1\,.
\label{gammakr}
\eeq

\noindent
Let us rewrite the growth rate in terms of a \emph{new characteristic frequency}, defined by: 
\beq
\sigma \;=\; \vert\eta_K\vert k_\alpha^2
\;=\; \left(\frac{\vert\eta_T - \eta_\alpha\vert}{\eta_\alpha}\right)\frac{1}{\tau_\alpha}
\;\geq\; 0\,.
\label{Sstar}
\eeq

\noindent 
Then (\ref{gammakr}) can be rewritten as:
\beq
\gamma \;=\; 
\pm\;\sigma\left(\frac{k}{k_\alpha}\right)^2
\left[\left(\frac{k}{k_\alpha}\right)^2 \,-\, 1\right]\,,\qquad\mbox{when}\qquad \vert k\vert \ll \frac{k_\alpha}{\sqrt{\sigma\tau_\alpha}}\,,
\label{gammakrs}
\eeq

\noindent
where the $\pm$ signs correspond to the cases of {\bf weak $\alpha$ fluctuations} ($\eta_K > 0$) and {\bf strong $\alpha$ fluctuations} 
($\eta_K < 0$), respectively. As Figure~1 shows, for the weak case $\gamma$ is negative 
for low wavenumbers and positive for high wavenumbers, and is exactly the
opposite for the strong case. Hence there is a high wavenumber dynamo for weak $\alpha$ fluctuations, and a low wavenumber dynamo for strong 
$\alpha$ fluctuations.\footnote{It should be noted that this behaviour is \emph{not} contradictory to the white--noise case: in the limit $\tau_\alpha \to 0$, both $k_\alpha \to \infty$ and $\sigma \to \infty$, and $\gamma \to 
\mp (\sigma/k_\alpha^2)k^2 = -\eta_K k^2\,$ which is in agreement with
the results for the white--noise case.} As noted earlier, the high wavenumber behaviour is likely an overestimate because the dissipative term, $\eta_T\bnabla^2\bfh$, was dropped in the FOSA equation~(\ref{FFE-fosa}).

\section{Growth rates of modes when $\tau_\alpha$ is non zero}

We consider one--dimensional propagating modes for the general case
when all the parameters $(\eta_\alpha, S, \bfV_{\!\!M}, \tau_\alpha)$
can be non zero. Below we derive the dispersion relation and study the growth rate function.  When the wavevector $\bfk = (0,0,k)$ points along the ``vertical'' ($\pm\ez$) directions, $\widetilde{\overline{H}}_3$ must be 
uniform and is of no interest for dynamo action. Hence we set $\widetilde{\overline{H}}_3 = 0$, and take $\widetilde{\overline{\bfH}}(k, t) = \widetilde{\overline{H}}_1(k, t)\ex \,+\, \widetilde{\overline{H}}_2(k, t)\ey\,$. The equation governing the time evolution of this large--scale magnetic field is obtained by setting $k_1 = 0$, $\,k_3=k\,$ and $\,\widetilde{\overline{H}}_3 =0\,$ in equation~(\ref{MFE-k2-0}):
\begin{eqnarray}
\frac{\partial \widetilde{\overline{\bfH}}}{\partial t} &\;=\;&
\left[\,S\widetilde{\overline{H}}_1\ey \,-\, \eta_K k^2 \widetilde{\overline{\bfH}} \,-\,\mathrm{i}\,kV_{M3} \widetilde{\overline{\bfH}}\,\right]
\left[\,1 \,+\, \mathrm{i}\,kV_{M3}\tau_{\alpha} \,-\,
\eta_{\alpha} k^2\tau_{\alpha}\,\right] 
\nonumber\\[1em] 
&&\qquad +\;\;S\left[\,\mathrm{i}\,kV_{M3}\tau_{\alpha} \,-\,
\eta_{\alpha} k^2\tau_{\alpha}\,\right] 
\widetilde{\overline{H}}_2\ex\,,
\label{MFEk_3}
\end{eqnarray}

\noindent
We seek modal solutions of the form 
\beq
\widetilde{\overline{\bfH}}(k, t) \;=\; \left[\widetilde{\overline{H}}_{01}(k)\ex \,+\, \widetilde{\overline{H}}_{02}(k)\ey\right]\,\exp{(\lambda t)}\,.
\label{modal}  
\eeq

\noindent
Substituting this in equation~(\ref{MFEk_3}) yields the following \emph{dispersion relation for the two $(\pm)$ modes}: 
\begin{eqnarray}
\lambda_{\pm} &\;=\;&
-\left[\eta_Kk^2 \,+\, \mathrm{i}\,kV_{M3}\right]  
\left[\,1 \,+\, \mathrm{i}\,kV_{M3}\tau_{\alpha} \,-\,
\eta_{\alpha} k^2\tau_{\alpha}\,\right] 
\nonumber\\[1em]
&&\qquad\quad \pm\;\vert S\vert\,\sqrt{
\left[\mathrm{i}kV_{M3}\tau_\alpha - \eta_\alpha k^2\tau_\alpha\right]
\left[\,1 \,+\, \mathrm{i}\,kV_{M3}\tau_{\alpha} \,-\,
\eta_{\alpha} k^2\tau_{\alpha}\,\right]\;}
\label{disprel}
\end{eqnarray}

\noindent
Of particular interest is the growth rate $\gamma = {\rm Re}\{\lambda\}\,$, because dynamo action corresponds to the case when $\gamma > 0\,$. From 
the dispersion relation~(\ref{disprel}) we have:
\begin{eqnarray}
\gamma_{\pm} \;=\; {\rm Re}\{\lambda_{\pm}\} &\;=\;& \eta_K k^2 \left(\,\eta_{\alpha} k^2 \tau_{\alpha} - 1\right) \;+\; \tau_{\alpha} (kV_{M3})^2
\;\pm\; \vert S\vert\left[\chi_R^2 \,+\, \chi_I^2\right]^{1/4}\,\cos{(\psi/2)}\,,
\nonumber\\[1em]
\mbox{where}\quad \tan{(\psi)} &\;=\;& \frac{\;\chi_I\;}{\;\chi_R\;}\,,
\nonumber\\[1em]
\chi_R &\;=\;& \eta_\alpha k^2\tau_\alpha \left(\,\eta_{\alpha} k^2 \tau_{\alpha} - 1\right) \;-\; \left(kV_{M3}\tau_{\alpha}\right)^2
\,, \nonumber\\[1em]
\chi_I &\;=\;& -kV_{M3}\tau_{\alpha} \left(\,2\eta_{\alpha} k^2 \tau_{\alpha} \,-\, 1\right) \,.
\label{gamma}
\end{eqnarray}

\noindent
\emph{Note that both $\lambda$ and $\gamma$ are linear in the rate of shear $S$, when the parameters $(\eta_K, \eta_{\alpha}, V_{M3}, \tau_\alpha)$ are independent of $S$}. Define the dimensionless quantities
\begin{eqnarray}
\Gamma_{\pm} &\;=\;& \gamma_{\pm}\tau_\alpha\,,\qquad\qquad \beta = \eta_\alpha k^2\tau_\alpha\,,\nonumber\\[1ex]
\varepsilon_S &\;=\;& S\tau_\alpha\,,\qquad\qquad
\varepsilon_K \;=\; \eta_Kk^2\tau_\alpha\,,\qquad\qquad
\varepsilon_M \;=\; kV_{M3}\tau_\alpha\,.
\label{dimless}
\end{eqnarray}

\noindent
Here $\Gamma_{\pm}$ are the dimensionless growth rates of the $\pm$ 
modes. $\beta = (k/k_\alpha)^2$ is a measure of the wavenumber $k$ 
in units of the characteristic wavenumber $k_\alpha$ defined in equation~(\ref{kalpha}), and can take any non--negative value. $\left(\varepsilon_S\,,\varepsilon_K\,, \varepsilon_M\right)$ must all be small by the three conditions in (\ref{cond}) under which the basic mean field equation~(\ref{MFE-k2-0}) is valid. There is just one constraint involving $\beta$ and $\varepsilon_K$, coming from $\beta + \varepsilon_K = \eta_Tk^2\tau_\alpha > 0\,$. Thus the parameter ranges are given by:
\begin{eqnarray}
&&0 \;\leq\; \beta \;<\; \infty\,, \qquad\qquad \beta \;+\; 
\varepsilon_K \;>\; 0\,,
\nonumber\\[1ex]
&&\vert\varepsilon_S\vert \;\ll\; 1\,,\qquad\qquad
\vert\varepsilon_K\vert \;\ll\; 1\,,\qquad\qquad
\vert\varepsilon_M\vert \;\ll\; 1\,,
\label{parrange}
\end{eqnarray}

\noindent
Multiplying equation~(\ref{gamma}) by $\tau_\alpha > 0\,$, we obtain 
the dimensionless growth rates,
\begin{eqnarray}
\Gamma_{\pm} &\;=\;& \varepsilon_K(\beta \,-\, 1) \;+\; \varepsilon_M^2
\;\pm\; \vert\varepsilon_S\vert\left[\beta^2(\beta - 1)^2 \,+\, \varepsilon_M^2\left(2\beta^2 - 2\beta + 1\right)\,+\, \varepsilon_M^4\right]^{1/4}\,\cos{(\psi/2)}\,,
\nonumber\\[1em]
&&\mbox{where}\qquad\tan{(\psi)} \;=\; 
\frac{\varepsilon_M(1 - 2 \beta)}
{\beta(\beta - 1) - \varepsilon_M^2}\,,
\label{Gamma}
\end{eqnarray}

\noindent
as a function of the 4 dimensionless parameters $\left(\beta\,, \varepsilon_S\,,\varepsilon_K\,, \varepsilon_M\right)$. Let $\Gamma_>$
be the larger of $\Gamma_+$ and $\Gamma_-$, and $\Gamma_<$
be the smaller of $\Gamma_+$ and $\Gamma_-$. Then, from equation~(\ref{Gamma}), we have:
\beq
\Gamma_{\stackrel{>}{<}} \;=\; \varepsilon_K(\beta \,-\, 1) \;+\; \varepsilon_M^2 \;\pm\; \vert\varepsilon_S\vert\left[\beta^2(\beta - 1)^2 \,+\, \varepsilon_M^2\left(2\beta^2 - 2\beta + 1\right)\,+\, \varepsilon_M^4\right]^{1/4}\,\vert\cos{(\psi/2)}\vert\,.
\label{gammagrls}
\eeq

\noindent
\emph{Note that $\Gamma_>$ increases, whereas $\Gamma_<$ decreases linearly with the shearing rate $S$, when the parameters $(\eta_K, \eta_{\alpha}, V_{M3}, \tau_\alpha)$ are independent of $S$.}

\subsection{The growth rate function $\Gamma_>$}

We now study the behaviour of $\Gamma_>\,$, given in equation~(\ref{gammagrls}), as a function of the 4 parameters $\left(\,\beta\,,\varepsilon_S\,,\varepsilon_K\,, \varepsilon_M\right)$ for the range 
of parameter values given in (\ref{parrange}). Of these, 
the three parameters $\left(\varepsilon_S\,,\varepsilon_K\,, \varepsilon_M\right)$ can be taken to be independently specified, taking positive 
and negative values, so long as their magnitudes are small. But $\beta\geq 0\,$ is subject to the constraint $\beta + \varepsilon_K > 0\,$. Therefore
we can rewrite the conditions of (\ref{parrange}) as:
\begin{eqnarray}
&&\vert\varepsilon_S\vert \;\ll\; 1\,,\qquad\qquad
\vert\varepsilon_K\vert \;\ll\; 1\,,\qquad\qquad
\vert\varepsilon_M\vert \;\ll\; 1\,,
\nonumber\\[1ex]
&& \mbox{For $\;\varepsilon_K \leq 0\,$, have $\;\vert\varepsilon_K\vert \;<\; \beta \;<\;
\infty\,$; $\qquad$ For $\;\varepsilon_K > 0\,$, have $\;0 \;\leq\; \beta \;<\;
\infty\,$.}
\label{parrange2}
\end{eqnarray}

\noindent
Taking advantage of the smallness of $\vert\varepsilon_M\vert$, the 
expression for $\Gamma_>$ in equations~(\ref{gammagrls}) can be simplified
by (i) noting that $\left(2\beta^2 - 2\beta + 1\right) \geq 1/2$ for any real $\beta\,$ implies that $\varepsilon_M^2\left(2\beta^2 - 2\beta + 1\right) \gg \varepsilon_M^4$, so the $\varepsilon_M^4$ term inside the $[\;\;]^{1/4}$ can be dropped; (ii) taking the limit $\varepsilon_M \to 0\,$ in $\vert\cos{(\psi/2)}\vert\,$. The two \emph{special cases}, $\beta =0$ and $\beta = 1\,$, are dealt with first. Then we take up the \emph{general case}, and derive a simple expression for $\Gamma_>$ for arbitrary positive 
$\beta$, so long as it is not too close to the special values $(0\,, 1)\,$.
\emph{Taken together, the two special cases and the general case enables us to write a simple, good approximation to $\Gamma_>$ as a function of the 4 parameters $\left(\,\beta\,,\varepsilon_S\,,\varepsilon_K\,, \varepsilon_M\right)$, all independently varying: the $\varepsilon$'s being small in 
magnitude, and $0<\beta<\infty\,$}.

\noindent
\emph{1. Case $\beta \,=\, 0\,$}:
 
From (\ref{parrange2}) we must
have $\varepsilon_K > 0\,$, whereas $\varepsilon_S$ and $\varepsilon_M$ can be of either sign. From (\ref{gammagrls}) we have $\tan(\psi) = -(1/\varepsilon_M) \to \pm\infty\;$ for small and negative/positive 
$\varepsilon_M\,$, which implies that $\vert\cos{(\psi/2)}\vert = 1/\sqrt{2}\,$. Hence:
\beq
\Gamma_> \;\;\simeq\;\; -\,\varepsilon_K \;+\; \varepsilon_M^2 \;+\; 
\frac{1}{\sqrt{2}}\vert\varepsilon_S\vert\,\vert\varepsilon_M\vert^{1/2}\,.
\label{betazero}
\eeq

\noindent
\emph{2. Case $\beta \,=\, 1\,$}: 

From (\ref{parrange2}) we can see that all the three small parameters,
$\left(\varepsilon_K\,, \varepsilon_S\,, \varepsilon_M\right)\,$, can be of either sign. From (\ref{gammagrls}) we have $\tan(\psi) = (1/\varepsilon_M) \to \pm\infty\;$ for small and positive/negative $\varepsilon_M\,$, which implies that $\vert\cos{(\psi/2)}\vert = 1/\sqrt{2}\,$. Hence: 
\beq
\Gamma_> \;\;\simeq\;\; \varepsilon_M^2 \;+\; 
\frac{1}{\sqrt{2}}\vert\varepsilon_S\vert\,\vert\varepsilon_M\vert^{1/2}
\,.
\label{betaone}
\eeq

\noindent
\emph{3. General Case: $\beta > 0\,$ but not too close to $0$ or $1\,$}:

Excluding two narrow strips around $0$ and $1\,$, we consider the 
parameter ranges: 
\begin{eqnarray}
&& \vert\varepsilon_S\vert \;\ll\; 1\,,\qquad\qquad
\vert\varepsilon_K\vert \;\ll\; 1\,,\qquad\qquad
\vert\varepsilon_M\vert \;\ll\; 1\,,
\nonumber\\[1ex]
\qquad\mbox{and}\qquad
&&\beta_{\rm min} \;\leq\; \beta \;<\; \left(1 \,-\, \beta_{\rm min}\right)\qquad\mbox{and}\qquad
\left(1 \,+\, \beta_{\rm min}\right) \;<\; \beta \;<\; \infty\,,
\nonumber\\[1ex]
\qquad\mbox{where}\qquad
&& {\rm Max}\{\vert\varepsilon_K\vert\,, \vert\varepsilon_M\vert\}\;\ll\; \beta_{\rm min} \;\ll\; 1\,.
\label{regfin}
\end{eqnarray}

\noindent
Note that this is not only compatible with conditions (\ref{parrange2}), but they are only slightly less general. The small parameters $\left(\varepsilon_K\,, \varepsilon_S\,, \varepsilon_M\right)\,$, can be of either sign. Since $\beta_{\rm min} \gg
\vert\varepsilon_M\vert\,$, we have: (i) In $\Gamma_>$ of (\ref{gammagrls}), the term $[\;\;]^{1/4} \simeq [\beta^2(\beta - 1)^2]^{1/4}
= \sqrt{\,\vert\beta(\beta - 1)\vert\,}\,$; (ii) $\tan(\psi) \simeq 
\varepsilon_M(1-2\beta)/\beta(\beta - 1) \to 0\;$ for small and positive/negative $\varepsilon_M\,$, which implies that $\vert\cos{(\psi/2)}\vert = 1\,$. Hence: 
\beq
\Gamma_> \;\;\simeq\;\; 
\varepsilon_K(\beta \,-\, 1) \;+\; 
\varepsilon_M^2 \;+\; 
\vert\varepsilon_S\vert\,\sqrt{\vert\beta(\beta \,-\, 1)\vert\;}
\,.
\label{betagen}
\eeq

\noindent
Our analysis of dynamo action below is based on the above three cases. 

\section{Dynamo action due to Kraichnan diffusivity, Moffatt drift and Shear}

\begin{figure}
\begin{center}
\includegraphics[width=\columnwidth]{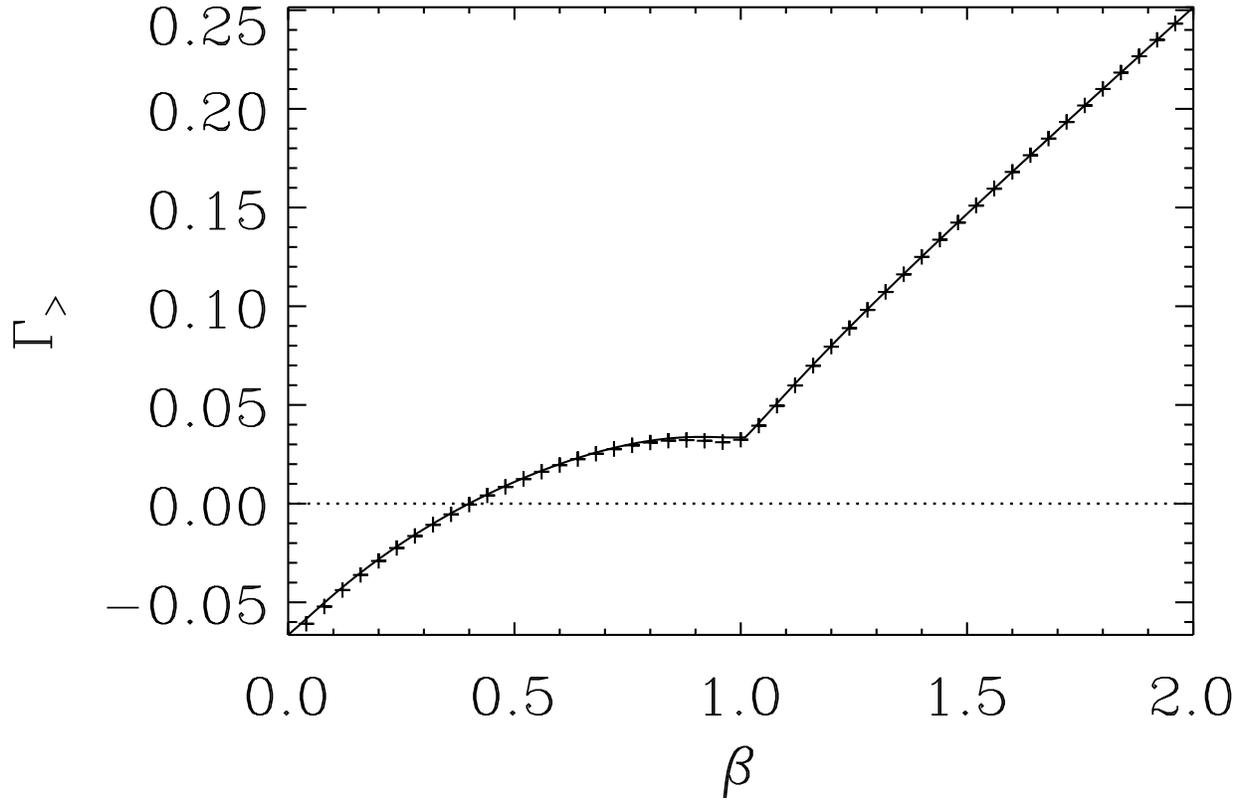}
\end{center}
\caption[]{Growth rate function $\Gamma_>$ plotted as a function of
$\beta$, for  $\varepsilon_K = 0.1$ , $\vert\varepsilon_M\vert = 0.1$
and $\vert\varepsilon_S\vert = 0.1$. The bold line corresponds to the 
exact expression of equation~(\ref{gammagrls}), and the `$+$' symbols correspond to the interpolation formula of equation~(\ref{betafin}).}
\label{Fig2}
\end{figure}

We are now ready to deal directly with the physical properties of dynamos driven by Kraichnan diffusivity, Moffatt drift and Shear. From the three cases considered above in (\ref{betazero})---(\ref{betagen}), we write down:  
\beq
\Gamma_> \;\;\simeq\;\;
\varepsilon_K(\beta \,-\, 1) \;+\; 
\varepsilon_M^2 \;+\; 
\vert\varepsilon_S\vert\left[\;\beta^2(\beta \,-\, 1)^2
\;+\; \frac{\varepsilon_M^2}{4}\;\right]^{1/4}
\,,
\label{betafin}
\eeq

\noindent
as a simple interpolation formula, \emph{valid for the independent variations of all 4 parameters}: as before the three $\varepsilon$'s are small, but can freely take positive and negative values; $\beta$ in 
now free to take any positive value in an unconstrained manner:
\beq
\vert\varepsilon_S\vert \;\ll\; 1\,,\qquad\qquad
\vert\varepsilon_K\vert \;\ll\; 1\,,\qquad\qquad
\vert\varepsilon_M\vert \;\ll\; 1\,, \qquad\qquad
0 \;<\; \beta \;<\; \infty\,.
\label{parrange3}
\eeq

\noindent
Figure~2 compares the general formula (\ref{gammagrls}) for the 
dimensionless growth rate $\Gamma_>\,$, with the interpolation formula
given in equation (\ref{betafin}) above; as may be seen it is an excellent fit to (\ref{gammagrls}). Note that, even though $\varepsilon_K >0\,$
(corresponding to weak $\alpha$ fluctuations), $\Gamma_>\,$ is positive
for a range of values of $\beta\,$, allowing for dynamo action.
 A general property of equation (\ref{betafin}) is that, for fixed $\beta$ and $\varepsilon_K\,$, the dimensionless growth rate $\Gamma_>$
is a monotonically increasing function of both $\vert\varepsilon_M\vert$
and $\vert\varepsilon_S\vert\,$. Hence we arrive at an important 
conclusion:

\begin{itemize}
\item[]
\emph{Both Moffatt drift and shear always act to increase the growth 
rate of traveling waves of the form, $\;\overline{\bfB}(X_3, \tau) \,=\, \left[\overline{B}_1\ex \,+\, \overline{B}_2\ey\right]\exp{\!\left({\rm i}kX_3 + \lambda \tau\right)}\,$, where $B_1$ and $B_2$ are arbitrary constants and $\lambda$ is a complex frequency}.
\end{itemize}

\noindent
We have already considered the combined effect of non zero $\tau_\alpha$
and Kraichnan diffusivity in \S~5.2. Below we present two examples
of dynamo action due to non zero $\tau_\alpha$ in conjunction with Moffatt drift and Shear.

\subsection{A Moffatt drift dynamo}

We consider the case of zero shear, $\varepsilon_S = 0$, but general
values of $\varepsilon_K$ and $\varepsilon_M$. This can be thought of as the Kraichnan--Moffatt problem, discussed in \S~2.1, generalized to include a non zero $\tau_\alpha\,$. From equation~(\ref{MFEk_3}) we see that, for $S=0$, the two components of $\widetilde{\overline{\bfH}}(k, t)$ evolve independently. The growth rate given by equation~(\ref{betafin}) is:
\beq
\Gamma_> \;\;\simeq\;\; 
\varepsilon_K(\beta \,-\, 1) \;+\; 
\varepsilon_M^2\,.
\label{no_shear}
\eeq

\noindent
The simplest cases are when one of $\varepsilon_K$ and $\varepsilon_M$ is zero: $\varepsilon_M = 0$ has already been discussed in \S~5.3, so let us consider the complementary case $\varepsilon_K = 0\,$. Then $\Gamma_> = \varepsilon_M^2$ is non--negative, so that the dimensional growth rate
\beq
\gamma_> \;=\; \left(V_{M3}^2\tau_\alpha\right)\,k^2\,,
\eeq

\noindent
is negatively diffusive, with negative--diffusion coefficient equal to 
$\left(V_{M3}^2\tau_\alpha\right)\,$. This growth is due to  
an extra term, ${\rm i}V_{M3}^2\tau_\alpha k\left(\ez\cross\widetilde{\overline{\bfH}}\right)$, in the EMF of equation~(\ref{EMF-taualp2}). \emph{Thus Moffat drift and a non zero $\alpha$--correlation time together can enable dynamo action}. This should be contrasted with the white--noise case where the Moffatt drift does not influence the growth rate.  
The remarkable thing is just that a constant drift velocity can enable a dynamo in the absence of any other generative effect. This is only possible if the relation between mean EMF and mean field has a memory effect. 
\citet{Rhe14} is another example where there is a drift velocity (a $\gamma$ effect), analogous to the Moffat drift in the case of the Roberts flow III. 
  
Now we go on to look at the more general problem.
When $\varepsilon_K$ and $\varepsilon_M$ are both non zero, then $\eta_K$ and $V_{M3}$ are also non zero; this gives rise to a \emph{new time scale} in the problem:
\beq
\tau_{*} \;=\; (\vert\eta_K\vert/V_{M3}^2) \;>\; 0\,.
\label{taustar}
\eeq

\noindent
Below we consider the behaviour of the dimensional growth rate $\gamma_>\,$, 
as a function of the wavenumber $k\,$, for weak and strong $\alpha$
fluctuations. As we will see, the nature of dynamo action depends on whether $\tau_\alpha$ is larger or smaller than $\tau_*\,$.

\begin{figure}
\begin{center}
\includegraphics[width=\columnwidth]{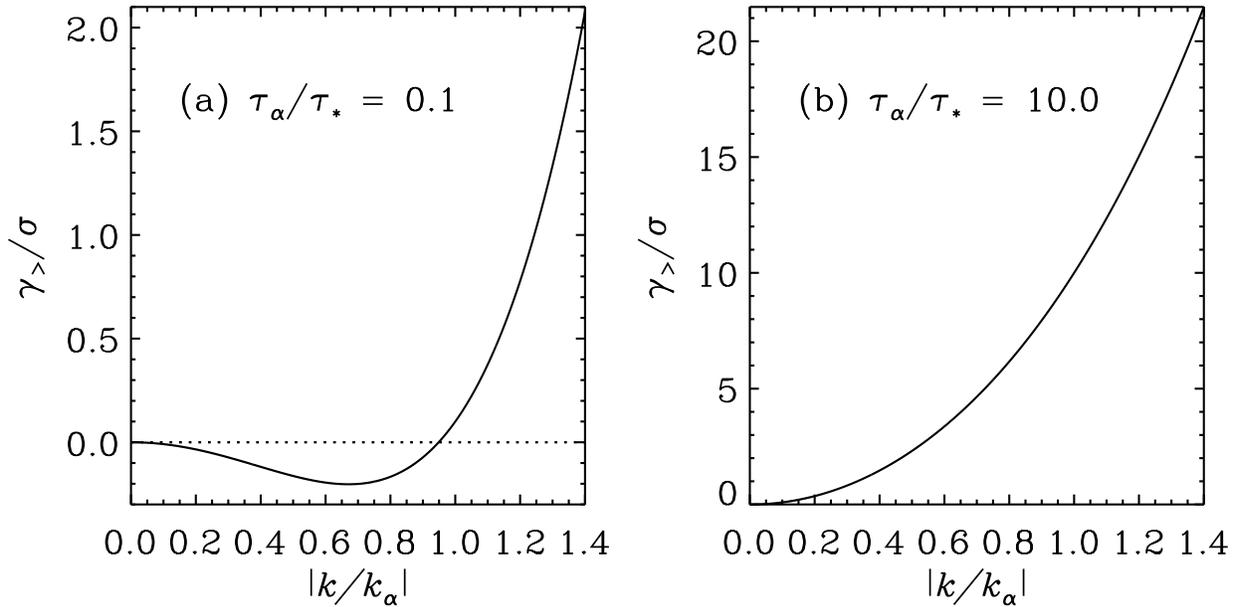}
\end{center}
\caption[]{Growth rate $\gamma_>/\sigma$ plotted as a function of
$\vert k/k_{\alpha}\vert$ for \emph{weak $\alpha$ fluctuations}, when 
$S=0\,$. Panels (a) and (b) correspond to the case when $\tau_{\alpha}/\tau_* = 0.1$ and $10.0$, respectively. Note that the growth rate is 
always positive for high wavenumbers.}
\label{Fig3}
\end{figure}

\begin{figure}
\begin{center}
\includegraphics[width=\columnwidth]{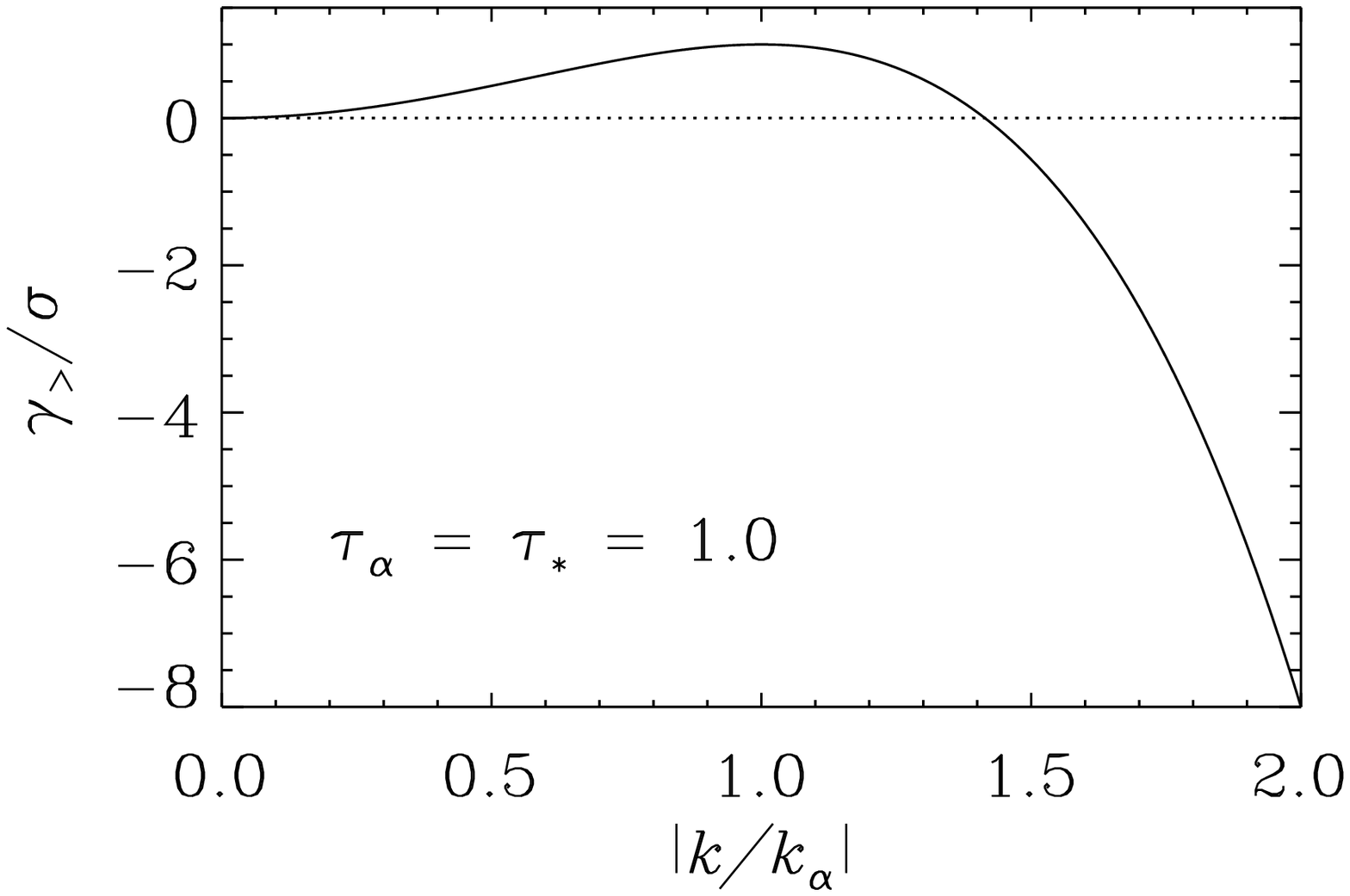}
\end{center}
\caption[]{Growth rate $\gamma_>/\sigma$ plotted as a function of
$\vert k/k_{\alpha}\vert$ for \emph{strong $\alpha$ fluctuations},
when $S=0\,$. Note that the growth rate is always positive at low wavenumbers, and negative for high enough wavenumbers.}
\label{Fig4}
\end{figure}
\bigskip

\medskip
\noindent
{\bf Weak $\alpha$ fluctuations:} This has $0<\eta_\alpha < \eta_T$, 
so that $\eta_K$ and $\varepsilon_K$ are both positive. The dimensional 
growth rate can be written as:
\beq
\gamma_> \;\;=\;\; \sigma\left\{\,
\left(\frac{k}{k_\alpha}\right)^4 \,+\, \left[\frac{\tau_\alpha}{\tau_*} \,-\, 1\right]
\left(\frac{k}{k_\alpha}\right)^2\right\}\,.
\eeq

\noindent
where $\sigma = \vert\eta_K\vert k_\alpha^2$ is the characteristic frequency
defined in equation~(\ref{Sstar}). High wavenumbers, $\vert k\vert > k_\alpha\,$,
always grow, with $\gamma_>$ being a positive and monotonically increasing
function of $\vert k\vert\,$. This behaviour is opposite, 
in a qualitative sense, to the white--noise case where weak $\alpha$
fluctuations always imply decaying large--scale fields.
When $\vert k\vert \gg k_\alpha\,$, the growth rate tends to the asymptotic form  $\gamma_> \,\to\, \sigma(k/k_\alpha)^4\,$ which
is independent of the Moffatt drift.  There are two cases
to study:

\begin{itemize}
\item[(a)]
$\underline{\mbox{Case}\;\tau_\alpha < \tau_*\,}$: For small enough wavenumbers $\vert k\vert < k_\alpha\sqrt{1 - \tau_\alpha/\tau_*}\,$, the growth rate is negative; for very small wavenumbers, there is positive diffusion with $\gamma_>\propto -k^2\,$. For large enough wavenumbers $\vert k\vert > k_\alpha\sqrt{1 - \tau_\alpha/\tau_*}\,$, the growth rate becomes positive and tends to being negatively super--diffusive $(\gamma_> \propto k^4$) for very large wavenumbers. The graph is given in Figure~3a.

\item[(b)]
$\underline{\mbox{Case}\;\tau_\alpha > \tau_*\,}$: The growth rate $\gamma_>$ is always positive for all  wavenumbers $k\,$.   
\emph{Therefore, $\tau_\alpha > \tau_*\,$ is a sufficient condition for dynamo action at all wavenumbers. This is equivalent to requiring that the Moffatt drift and the $\alpha$--correlation time be large enough such that}:
\beq
V_{M3}^2\tau_\alpha \,\;>\;\, \eta_K \,\;>\;\, 0\,.
\eeq

\noindent
As can be seen from Figure~3b, the growth is negatively diffusive 
($\gamma_>\propto k^2$) when $\vert k\vert \ll k_\alpha\,$, and 
negatively super--diffusive $(\gamma_> \propto k^4$) when $\vert k\vert \gg k_\alpha\,$. 
\end{itemize}

\medskip
\noindent
{\bf Strong $\alpha$ fluctuations:} This has $0<\eta_T < \eta_\alpha\,$, 
so that $\eta_K$ and $\varepsilon_K$ are both negative. The dimensional 
growth rate is:
\beq
\gamma_> \;\;=\;\; \sigma\left\{\,
-\left(\frac{k}{k_\alpha}\right)^4 \,+\, \left[\frac{\tau_\alpha}{\tau_*} \,+\, 1\right]
\left(\frac{k}{k_\alpha}\right)^2\right\}\,.
\eeq

\noindent
As shown in Figure~4, the growth rate $\gamma_>$ rises from zero at $\vert k\vert = 0$, to a maximum positive value,
\beq
\gamma_m \;=\; \frac{\sigma}{4}\left[\,
1 \,+\, \frac{\tau_\alpha}{\tau_*}\,\right]^2\,,
\qquad \mbox{at}\quad\vert k\vert \;=\; k_m \;=\; 
\frac{k_\alpha}{\sqrt{2}}\left[\,
1 \,+\, \frac{\tau_\alpha}{\tau_*}\,\right]^{1/2}\,.
\label{gammam} 
\eeq

\noindent
Then $\gamma_>$ decreases monotonically, turning negative for $\vert k\vert > k_\alpha\sqrt{1 + \tau_\alpha/\tau_*}\,$. For much larger $\vert k\vert$ the growth rate tends to the asymptotic form $\gamma_> \,\to\, -\sigma(k/k_\alpha)^4\,$ which is independent of the  Moffatt drift. Note that this behaviour is quite opposite to the white--noise case where strong $\alpha$ fluctuations always imply growing large--scale fields; but it nevertheless approaches smoothly the white--noise case for $\tau_\alpha \to 0\,$.

\subsection{A Shear dynamo}

We consider the case of zero Moffatt drift, $\varepsilon_M = 0$, but general
values of $\varepsilon_K$ and $\varepsilon_S$. The growth rate is:
\beq
\Gamma_> \;\;\simeq\;\; 
\varepsilon_K(\beta \,-\, 1) \;+\; 
\vert\varepsilon_S\vert\sqrt{\vert\beta(\beta \,-\, 1)\vert\,}\,,
\label{no_moff}
\eeq

\noindent
As earlier we will look at the properties of the dimensional growth rate
as a function of the wavenumber. 

\noindent
{\bf Weak $\alpha$ fluctuations:} This has $0<\eta_\alpha < \eta_T$, 
so that $\eta_K$ and $\varepsilon_K$ are both positive. From (\ref{no_moff})
and (\ref{Sstar}) the dimensional growth rate is:
\begin{eqnarray}
\gamma_> &\;=\;& \sigma
\left\vert\frac{k}{k_\alpha}\right\vert\,\sqrt{1 \,-\, \left\vert\frac{k}{k_\alpha}\right\vert^2\,}
\left[\,
\frac{\vert S\vert}{\sigma} \,\;-\;\, \left\vert\frac{k}{k_\alpha}\right\vert\,\sqrt{1 \,-\, \left\vert\frac{k}{k_\alpha}\right\vert^2\,}\,\right]\,,\quad
\mbox{when}\quad \vert k\vert < k_\alpha\,;
\nonumber\\[3ex]
&\;=\;& \sigma
\left\vert\frac{k}{k_\alpha}\right\vert\,\sqrt{\left\vert\frac{k}{k_\alpha}\right\vert^2 \,-\, 1\,}
\left[\,
\frac{\vert S\vert}{\sigma} \,\;+\;\, \left\vert\frac{k}{k_\alpha}\right\vert\,\sqrt{\left\vert\frac{k}{k_\alpha}\right\vert^2 \,-\, 1\,}\,\right]\,,\quad
\mbox{when}\quad \vert k\vert > k_\alpha\,.
\label{gammaweak}
\end{eqnarray}

\noindent
High wavenumbers, $\vert k\vert > k_\alpha\,$, always grow, with 
$\gamma_>$ being a positive and monotonically increasing function of $\vert k\vert\,$. This behaviour is opposite, in a qualitative sense, to the white--noise case where weak fluctuations decay. When $\vert k\vert \gg k_\alpha\,$, the growth rate tends to the asymptotic form  $\gamma_> \,\to\, \sigma(k/k_\alpha)^4\,$. For low wavenumbers, 
$\vert k\vert < k_\alpha\,$, the growth rate is positive for all wavenumbers only when $(\vert S\vert/\sigma)$ is larger than the maximum value of $(\vert k\vert/k_\alpha)
\sqrt{1-(\vert k\vert/k_\alpha)^2}\,$, which is equal to $1/2\,$.  
\emph{Therefore a sufficient condition for dynamo action at all 
wavenumbers is that $\vert S\vert > \sigma/2\,$. This is equivalent to 
requiring that the shearing rate and $\alpha$--correlation time 
be large enough such that}:
\beq
\vert S\vert\tau_\alpha \,\;>\;\, \frac{1}{2}\,\frac{\eta_K}{\eta_\alpha}
\,\;>\;\, 0\,.
\label{Slarge}
\eeq

\noindent
When the shearing rate is not large enough, $\vert S\vert < \sigma/2\,$, 
then at low wavenumbers, the growth rate is negative in a range of wavenumbers centered around $\vert k\vert = (k_\alpha/\sqrt{2})\,$. 
Figure~5 shows the growth rate as a function of wavenumber, for the 
two cases of weak and strong shear.

\begin{figure}
\begin{center}
\includegraphics[width=\columnwidth]{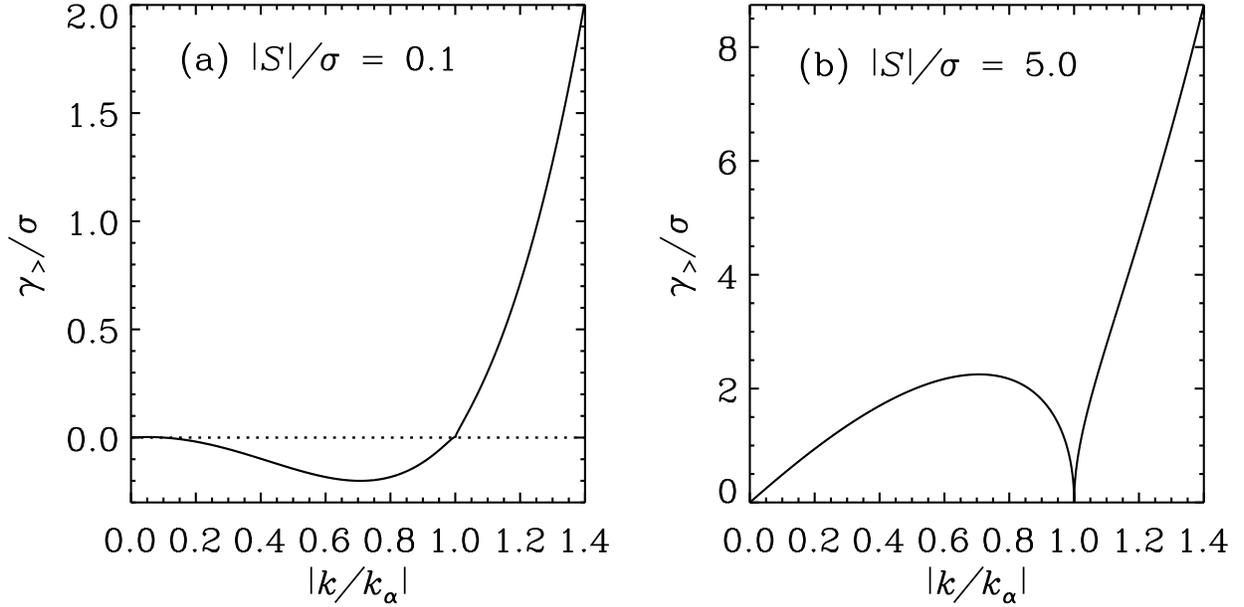}
\end{center}
\caption[]{Growth rate $\gamma_>/\sigma$ plotted as a function of
$\vert k/k_{\alpha}\vert$ for \emph{weak $\alpha$ fluctuations}, when 
$V_{M3}=0\,$. Panels (a) and (b) correspond to the case when $\vert S\vert/\sigma = 0.1$ and $5.0$, respectively. Note that the growth rate is always positive for 
high wavenumbers.}
\label{Fig5}
\end{figure}

\begin{figure}
\begin{center}
\includegraphics[width=\columnwidth]{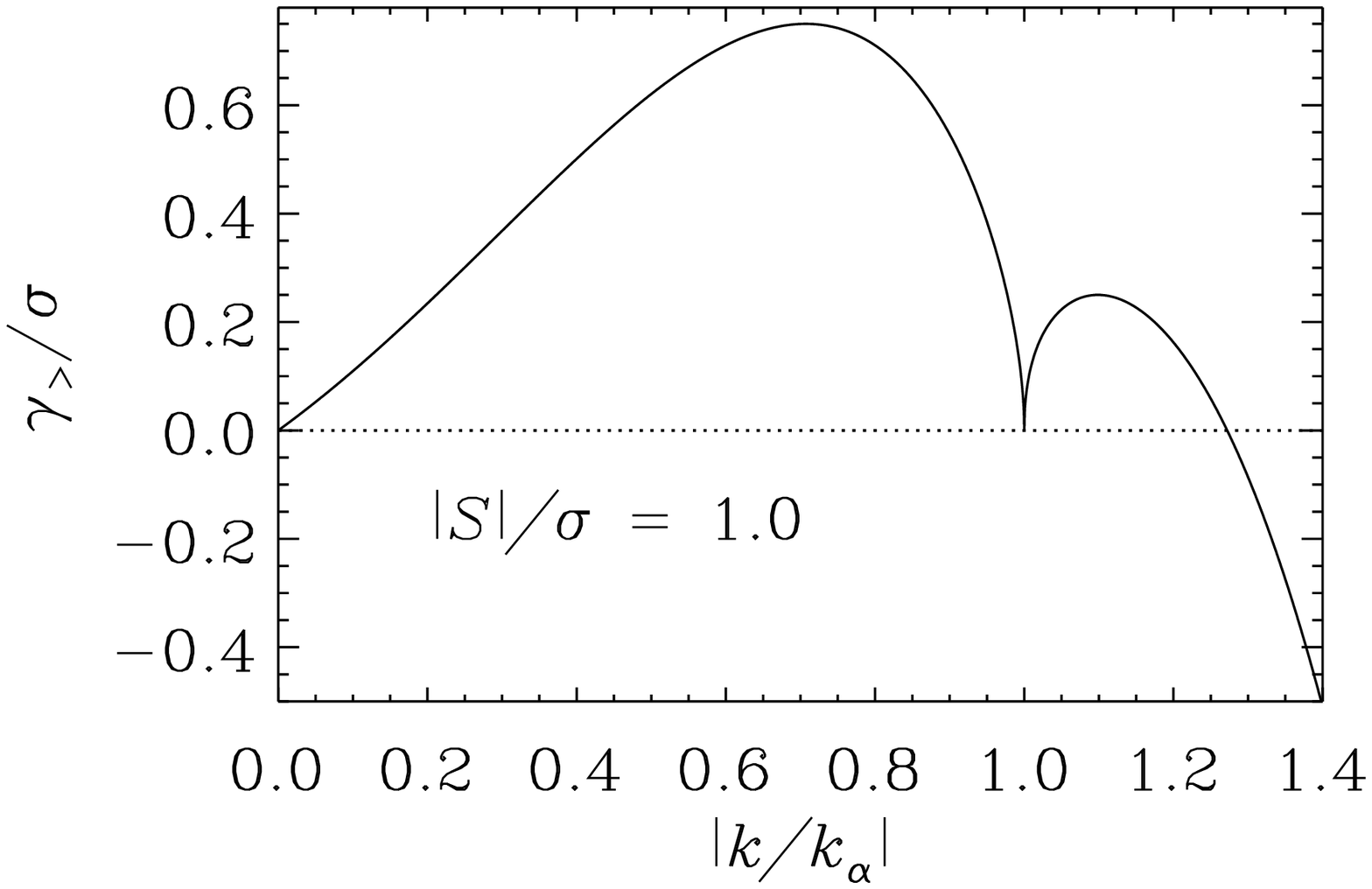}
\end{center}
\caption[]{Growth rate $\gamma_>/\sigma$ plotted as a function of
$\vert k/k_{\alpha}\vert$ for \emph{strong $\alpha$ fluctuations},
when $V_{M3}=0\,$. Note that the growth rate is always positive at low wavenumbers and negative for high enough wavenumbers.}
\label{Fig6}
\end{figure}

\noindent
{\bf Strong $\alpha$ fluctuations:} This has $0<\eta_T < \eta_\alpha\,$, 
so that $\eta_K$ and $\varepsilon_K$ are both negative. From (\ref{no_moff})
and (\ref{Sstar}) the dimensional growth rate is:
\begin{eqnarray}
\gamma_> &\;=\;& \sigma
\left\vert\frac{k}{k_\alpha}\right\vert\,\sqrt{1 \,-\, \left\vert\frac{k}{k_\alpha}\right\vert^2\,}
\left[\,
\frac{\vert S\vert}{\sigma} \,\;+\;\, \left\vert\frac{k}{k_\alpha}\right\vert\,\sqrt{1 \,-\, \left\vert\frac{k}{k_\alpha}\right\vert^2\,}\,\right]\,,\quad
\mbox{when}\quad \vert k\vert < k_\alpha\,;
\nonumber\\[1em]
&\;=\;&  \sigma
\left\vert\frac{k}{k_\alpha}\right\vert\,\sqrt{\left\vert\frac{k}{k_\alpha}\right\vert^2 \,-\, 1\,}
\left[\,
\frac{\vert S\vert}{\sigma} \,\;-\;\, \left\vert\frac{k}{k_\alpha}\right\vert\,\sqrt{\left\vert\frac{k}{k_\alpha}\right\vert^2 \,-\, 1\,}\,\right]\,,\quad
\mbox{when}\quad \vert k\vert > k_\alpha\,.
\label{gammastrong}
\end{eqnarray}

\noindent
Figure~6 shows the growth rate as a function of the wavenumber.
$\gamma_>$ rises from zero at $\vert k\vert = 0\,$, to a maximum positive value,
\beq
\gamma_{m1} \;=\; \frac{\vert S\vert}{2} \;+\; \frac{\sigma}{4}\,,
\qquad \mbox{at}\quad\vert k\vert \;=\; k_{m1} \;=\; 
\frac{k_\alpha}{\sqrt{2}}\,,
\label{gammam1}
\eeq

\noindent
and then decreases to zero at $\vert k\vert = k_\alpha\,$. For high 
wavenumbers, $\gamma_>$ again rises again from zero to a second maximum
positive value,
\beq
\gamma_{m2} \;=\; \frac{S^2}{4\sigma}\,,
\qquad \mbox{at}\quad\vert k\vert \;=\; k_{m2} \;=\; 
\frac{k_\alpha}{\sqrt{2}}\left[\,
1 \,+\, \sqrt{1 + \frac{S^2}{\sigma^2}\,}\,\right]^{1/2}\,, 
\label{gammam2}
\eeq

\noindent
and then decreases monotonically, turning negative for
$\vert k\vert > (k_\alpha/\sqrt{2})\left[1+\sqrt{1 + 4S^2/\sigma^2}\right]^{1/2}\,$. When $\vert k\vert \gg k_*\,$, the growth rate tends to the asymptotic form $\gamma_> \,\to\, -\sigma(k/k_\alpha)^4\,$. Note that this behaviour is quite opposite to the white--noise case where strong $\alpha$ fluctuations always imply growing large--scale fields; but it nevertheless approaches smoothly the white--noise case for $\tau_\alpha \to 0\,$.

\section{Conclusions}

We have presented a model of large--scale dynamo action in a linear shear flow that has stochastic, zero--mean fluctuations of the $\alpha$ parameter. 
This is based on a generalization of the Kraichnan--Moffatt (KM) model \citep{Kra76, Mof78}, to include a background linear shear and a non zero 
$\alpha$--correlation time. Our principal result is a linear integro--differential equation for the large--scale magnetic field, which is 
non--perturbative in both the shear ($S$) and the $\alpha$--correlation time
($\tau_\alpha$). 

An immediate application is to the case of \emph{white--noise $\alpha$ fluctuations} for which $\tau_\alpha = 0\,$. The electromotive force (EMF) turned out to be identical in form to the KM model without shear; this is because white--noise carries no memory whereas shear needs time to act. With no further approximation, the integro--differential equation was reduced to a partial differential equation (PDE), where shear entered it only through the background flow. We have presented a full solution of the initial value problem, and discussed the role of Kraichnan diffusivity ($\eta_K$), Moffatt drift ($\bfV_{\!\!M}$) and shear ($S$). Some salient results are (a) $\bfV_{\!\!M}\,$ contributes only to the phase and does not influence the growth of the field, which is qualitatively similar to the KM model with an additional $S$--dependent term; (b) The growth of the field depends on both $\eta_K$ and $S\,$; (c) However, the necessary condition for dynamo action is identical to the KM model (i.e. the $\alpha$ fluctuations must be strong, $\eta_K < 0\,$); then modes of all wavenumbers grow by shear--modified negative diffusion with the highest wavenumbers growing fastest. 

To explore memory effects on dynamo action it is necessary to let $\tau_\alpha$ be non zero. The integro--differential equation is difficult to analyze, hence we restricted attention to a simpler yet physically meaningful case where it reduces to a PDE; i.e. for slowly varying ``axisymmetric'' large--scale fields, small $\tau_\alpha$, and moderate values of $(\eta_K, \bfV_{\!\!M}, S)$ --- the precise conditions were stated in terms of three dimensionless small parameters. Working perturbatively, we derived an expression for the EMF that is accurate to first order in $\tau_\alpha\,$, and obtained the PDE governing the evolution of the large--scale magnetic field. We solved for exponential propagating modes with wavenumber $\bfk = k\ez$, and derived a dispersion relation. From this we obtained a formula for the growth rate, $\gamma_> = f + \vert S\vert g\,$, where $f$ and $g$ are explicitly given functions of  $(\eta_T\,, \eta_K\,, V_{M3}\,, \tau_\alpha\,, k)$. Therefore, if the parameters $(\eta_T\,, \eta_K\,, V_{M3}\,, \tau_\alpha)$ are independent of shear, then $\gamma_>$ is linear in $\vert S\vert$. The expression for $\gamma_>\,$ is somewhat complicated, so we simplified it to an interpolation formula that allowed us to explore dynamo action in some detail.  Here we recall some of the salient results:
\begin{itemize}
\item[1.] There can be dynamo action for even weak $\alpha$ fluctuations,
(i.e. positive Kraichnan diffusivity) when $\tau_\alpha\neq 0$. 

\item[2.]
Both Moffatt drift and Shear contribute to increasing the growth rate
at any wavenumber.

\item[3.] 
{\bf Moffatt drift dynamo:} There can be dynamo action when both $V_{M3}$ and $\tau_\alpha$ are non zero, but $S=0$. This is different from the 
KM model in which the Moffatt drift contributes only to the phase of 
the large--scale field and does not influence dynamo action. For the simplest case when $\eta_K=0$, the growth rate $\gamma_> = \left(V_{M3}^2\tau_\alpha\right)\,k^2$ is negatively diffusive, with negative--diffusion coefficient equal to $\left(V_{M3}^2\tau_\alpha\right)\,$; this growth is driven by an extra term, ${\rm i}V_{M3}^2\tau_\alpha k\left(\ez\cross\widetilde{\overline{\bfH}}\right)$, in the EMF of equation~(\ref{EMF-taualp2}). 
For more general values of $\eta_K$, we recall two results: 
For weak $\alpha$ fluctuations, $V_{M3}^2\tau_\alpha > \eta_K$ is a 
sufficient condition for dynamo action for all $k$. For strong 
$\alpha$ fluctuations, the growth rate as a function of the wavenumber
has a single positive maximum value, and turns negative for large 
$\vert k\vert$.

\item[4.]
{\bf Shear dynamo:} There can be dynamo action when both $S$ and $\tau_\alpha$ are non zero, but $V_{M3} = 0$. For weak $\alpha$ fluctuations, 
$\vert S\vert \tau_\alpha > \eta_K/(2\eta_\alpha)\,$ is a sufficient condition for dynamo action for all $k$. For strong $\alpha$ fluctuations, the growth rate as a function of the wavenumber has two positive maxima, 
and turns negative for large $\vert k\vert$.

\item[]
\emph{Cautionary note}: In the two types of dynamos discussed above, 
it is likely that the growth of high $\vert k\vert$ modes have been overestimated, because the dissipative term was dropped in the FOSA equation~(\ref{FFE-fosa}) for the fluctuating field.
\end{itemize}

Our model is based on the dynamo equation~(\ref{totmf_ensavd_expl})
for the meso--scale magnetic field. It is a minimal extension
of the KM dynamo equation~(\ref{DynEqn}) and inherits certain simplifying assumptions, such as the isotropy and locality of the transport coefficients $\eta_t$ and $\alpha$. Relaxation of these assumptions can, by itself, lead to dynamo action at the meso--scale level. Isotropy can be relaxed by making $\eta_t$ and $\alpha$ tensorial; see e.g. \citet{BRRK08, Dev13, SJ13, Rhe14} for the extraction of these transport coefficients from numerical simulations using the test--field method. Relaxing locality in space implies a $\bfk$--dependence of the Fourier--transformed transport coefficients, which we denote here by $\tilde{\eta_t}$ and $\tilde{\alpha}$; \citet{Dev13} have demonstrated dynamo action due to $\tilde{\eta_t}$ turning negative at low $k$ and overcoming molecular diffusivity. Relaxing locality in time \citep{HB09} makes $\tilde{\eta_t}$ and $\tilde{\alpha}$ dependent on $\omega$, which then implies memory effects; \citet{Rhe14} demonstrate dynamo action due to delayed transport with a tensorial $\tilde{\alpha}$. 

The other major assumption is the minimal role given to shear in our model; $\eta_t$ has been assumed constant, and the $\alpha$ fluctuations have a Galilean--invariant 2--point correlation function of a factored form; the time correlation part is ${\cal D}(t)$, and the spatial correlation part is ${\cal A}(\bfR)$, where $\bfR = \bfx-\bfx'+St'(x_1-x_1')\ey$. No restriction is placed on the value of $\eta_t$, or on the forms of the functions ${\cal D}(t)$ and $\bfA(\bfR)$: they could be either dependent on shear or not; it is just that our model does not specify this. In a more general model shear is likely to play a deeper role, by influencing the very nature of the underlying turbulence. Then we can no longer expect $\tilde{\eta_t}$ and $\tilde{\alpha}$ to be so simply behaved; they can be general tensorial functions of the many variables ($\bfk,\omega, S)$.  In this case, it is only natural to expect that the nature of dynamo action (criteria, wavenumber behaviour of the growth rate etc) may be different from items 1--4 above. Hence, particular results are less important than the formalism used to treat shear and $\alpha$--correlation time non--perturbatively. Can this formalism be extended to obtain an integro--differential equation for the large--scale magnetic field, corresponding to (\ref{MFE-fou}) and (\ref{EMF-etav}), in the more general case? Even when the answer is in the affirmative, one is faced with the task of choosing functional forms and fluctuation statistics for tensorial, many--variable functions. It is here that the test--field method may serve a very useful purpose, by extracting transport coefficients from numerical simulations. Once we have plausible functional forms, there are still two more tasks at hand: (a) the study of the dynamo modes of the model
and comparison with numerical simulations; (b) addressing the very nature of the underlying turbulence in a shear flow that gives rise to the transport coefficients.

\section*{Acknowledgments}
We are grateful to the referee Matthias Rheinhardt for useful comments and 
discussions, that stimulated us to undertake major revisions of our work.

\label{lastpage}

\end{document}